\documentclass[english,12pt,oneside]{amsproc}
\usepackage[utf8]{inputenc}
\usepackage[english]{babel}
\usepackage{a4wide}
\usepackage{amsthm}
\usepackage{graphics}
\usepackage{amsfonts, amssymb, amscd, amsmath}
\usepackage{latexsym}
\usepackage{color}
\usepackage{scalerel}
\usepackage{hyperref}

\newcommand{\bbrl}{\scalerel*{\{}{\sum}}
\newcommand{\bbrr}{\scalerel*{\}}{\sum}}
\newcommand{\ca}[1]{\mathcal{#1}}
\newcommand{\dd}{\partial}
\newcommand{\Ro}{\mathbb{R}}
\newcommand{\Qo}{\mathbb{Q}}
\newcommand{\tb}{t^{\birth}}
\newcommand{\td}{t^{\death}}

\newcounter{stmcounter}[section]
\newcounter{thcounter}

\numberwithin{equation}{section}

\theoremstyle{plain}

\usepackage{hyperref}
\DeclareMathOperator{\Ker}{Ker}
\DeclareMathOperator{\Imm}{Im}

\DeclareMathOperator{\supp}{supp}

\DeclareMathOperator{\birth}{birth}
\DeclareMathOperator{\death}{death}
\DeclareMathOperator{\Dowk}{Dowk}

\DeclareMathOperator{\dist}{dist}

\newtheorem{thm}[thcounter]{Theorem}

\newtheorem{prop}[stmcounter]{Proposition}

\theoremstyle{remark}

\newtheorem{rem}[stmcounter]{Remark}
\newtheorem{con}[stmcounter]{Construction}

\theoremstyle{definition}
\newtheorem{defin}[stmcounter]{Definition}

\begin{document}
	\title{Topology of cognitive maps}
	\author{K. Sorokin*, A.Ayzenberg, K. Anokhin,  R.Drynkin, V.Sotskov,  A. Zaitsew, M. Beketov}
	\date{Summer 2020 - Summer 2022}
	\thanks{The article was prepared within the framework of the HSE University Basic Research Program.}

	\maketitle
	
* ksorokin@hse.ru

\section*{Highlights}

--- A method for filtering out the place cells excluding the ''fake'' activations of different neurons is proposed. This allows collecting more data on the activity of place cells at the center of arena (which is often an issue). 
\newline
--- A method for constructing a simplicial complex on time series of neural groups' activities, retrieving the topological information about the arena (the method is applicable for studying the dynamics of different neural groups, specialized in their own way). 
\newline
--- A method for multidimensional scaling of nearest neighbors graphs obtained from a point cloud of neural groups' activities is applied to retrieve the topological information about the arena. 
\newline
--- A theoretical apparatus allowing to connect the above mentioned methods to each other in a mathematically adequate way is proposed and applied to our data.

\section*{Abstract}
In present paper we discuss several approaches to reconstructing the topology of the physical space from neural activity data of CA1 fields in mice hippocampus, in particular, having Cognitome theory of brain function \cite{KV} in mind. In our experiments, animals were placed in different new environments and discovered these moving freely while their physical and neural activity was recorded. We test possible approaches to identifying place cell groups out of the observed CA1 neurons. We also test and discuss various methods of dimension reduction and topology reconstruction. In particular, two main strategies we focus on are the Nerve theorem and point cloud-based methods. Conclusions on the results of reconstruction are supported with illustrations and mathematical background which is also briefly discussed.

\section*{Introduction}

The discovery of place cells was made in the fundamental work of O'Keefe and Dostrovsky ~\cite{OKeefe}. These are certain cells in the field CA1 of the hippocampal formation (HF) of mammalian brain reacting on animal's position in physical space. This led to a hypothesis that the space is encoded in memory by the place cells. Their existence was confirmed for mice \cite{Moser_mice}, rats ~\cite{OKeefe}, rabbits \cite{schvirkov}, bats \cite{Ulanovsky}, monkeys ~\cite{Ono} and humans ~\cite{Ekstrom}. This encoding of the physical space is referred to as the cognitive map. Conjecturally, the cognitive map can be decoded from neural cells' activity using various mathematical methods. 

The common experiment design in this research area is as follows. The animal is placed in a certain environment (arena) where it moves freely. As a camera records the position of an animal, the neural activity in its HF is being registered by one of the standard tracking methods: calcium imaging ~\cite{Cannell}, in our case. Then, among all the recorded neurons, certain cells are chosen, that correlate most with the animal's position in the arena. With this done, the problem is: can the topological properties of the environment be reconstructed from the time series of the chosen neurons' activity? And which approach is best suited for this problem?

\rem Present work is the first part of a planned series of papers on our approaches to using topological data analysis (TDA) techniques to reconstruct some basic cognitome features.   

One of the main novelties we propose in this paper is making a step towards focusing on system determination of neural groups, extracting a particular part of cognitive hypernetworks and studying their activity. System determination and functional groups were proposed more than half a century ago and developed in detail by Schvirkov \cite{schvirkov}. He demonstrated the intersections of neural functional systems in rabbits' visual cortex and the presence of equally determined systems within various other brain regions. 

\section*{Topology reconstruction}

\subsection{Basic notions}
Recall that  a simplicial complex is, informally, a generalization of a graph to a higher dimension, which often can be used to model the topology of some real data. Formally, a simplicial complex $K$ on a set $V$ is a collection of subsets of $V$ such that (1) for any $i\in V$ the singleton $\{i\}$ lies in $K$, (2) if $I\in K$ and $J\subset I$ then $J\in K$. Elements of $V$ are called vertices, elements of $K$ are called simplices. Any simplicial complex can be given a geometric interpretation in which vertices are represented by points, two-element subsets $\{i_1,i_2\}\in K$ are represented by edges between vertices $i_1$ and $i_2$, 3-element subsets $\{i_1,i_2,i_3\}\in K$ are represented by triangles spanned by vertices $i_1,i_2,i_3$, etc. This interpretation represents a compact topological space called the geometric realization of $K$. The dimension of a simplex $I\in K$ is, by definition, the number $\dim I=|I|-1$, one less than the number of its vertices.

We also recall the definition of simplicial homology and Betti numbers of $K$. Fix a base field $F$ (usually $F$ is either the field $\Qo$ of rational numbers or the field $F_2$ of two elements). Let $C_j(K)=F\langle I\in K\mid \dim I=j\rangle=\bbrl\sum_{I\in K, \dim I=j}a_II\mid a_I\in F\bbrr$ be the vector space over $F$ formally generated by $j$-dimensional simplices of $K$. Formally set $C_{-1}(K)=0$. Define the linear map (homomorphism) $\dd\colon C_j(K)\to C_{j-1}(K)$ called the simplicial differential (or the boundary map), by setting its values on the basis element $I=\{i_0,i_1,\ldots,i_j\}\in K$:
$$
\dd I=\sum_{s=0}^{j}(-1)^s\{i_0,\ldots,\widehat{i_s},\ldots,i_j\}\in C_{j-1}(K)
$$
(where $\widehat{i_s}$ means that the vertex $i_s$ was dropped from $\{i_0,\ldots,i_s,\ldots,i_j\}$)
and extending by linearity. It holds true that $\dd(\dd I)=0$ for any simplex $I\in K$. The kernels and images of the corresponding boundary maps: $Z_j(K) = \Ker(\dd\colon C_j(K)\to C_{j-1}(K))$ and $B_j(K) = \Imm(\dd\colon C_{j+1}(K)\to C_{j}(K))$ are vector spaces called the space of $j$-dimensional cycles and the space of $j$-dimensional boundaries of $K$ respectively. One obtains the following inclusions: $B_j(K)\subseteq Z_j(K)\subseteq C_j(K)$. The quotient $H_j(K)=Z_j(K)/B_j(K)$ forms a vector space called the $j$-th homology group of $K$. Its dimension is called the $j$-th Betti number of $K$:
\[
\beta_j(K)=\dim_FH_j(K).
\]
It is known that $\beta_0(K)$ equals the number of connected components of $K$, while $\beta_{j>0}(K)$ allows to analyze higher-dimensional topological features of $K$. Informally, $\beta_j(K)$ counts ``the number of $j$-dimensional holes in $K$''. If $K$ is a graph, the number $\beta_1(K)$ is also called the cyclomatic number of $K$. It counts the number of independent cycles of a graph. For example, any tree has $\beta_1=0$ while the figure eight has $\beta_1=2$, since it has two ``holes'' (two independent cycles). The notions of homology and Betti numbers can be extended to more general topological spaces, via different (always equivalent) homology theories like singular homology or (dual) \v{C}ech cohomology \cite{Bott}.

\subsection{Applying topological methods to real data}

\begin{figure}[!htb]
\centering
\includegraphics[width=0.9\linewidth]{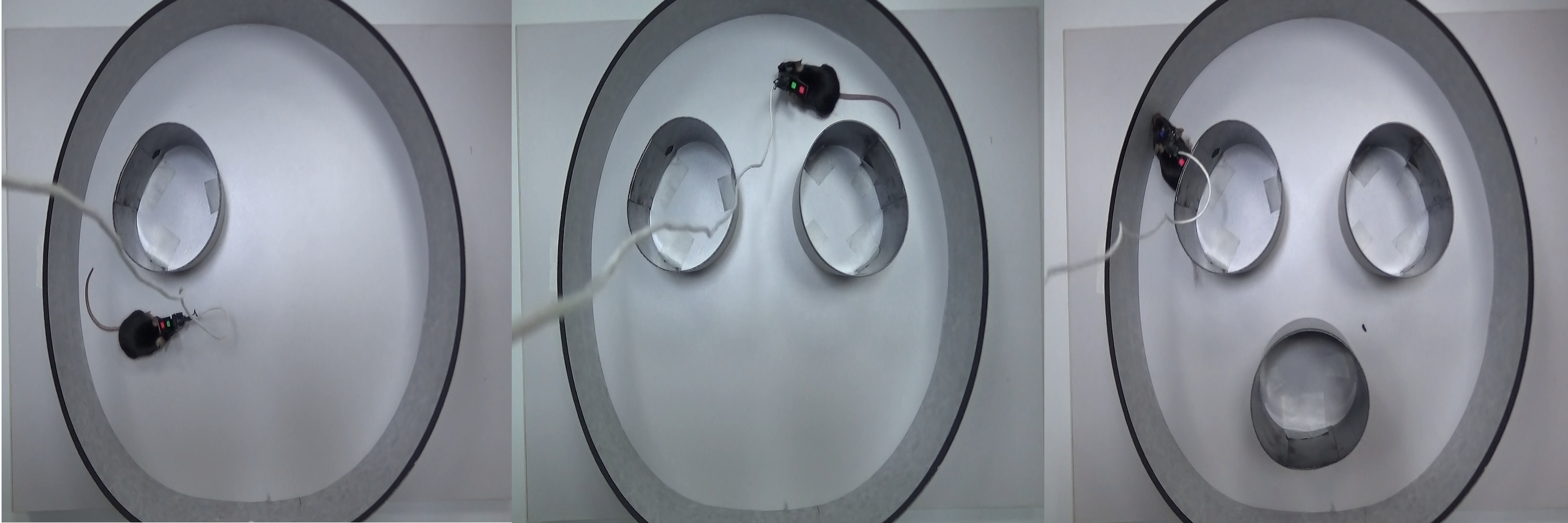}
\caption{Three arenas with different topologies ($ \beta_1 = 1, 2, 3$ correspondingly)} \label{fig:three arenas with different topologies}
\end{figure}

Three arenas with different topologies constructed by putting non-transient obstacles were prepared (Fig.\,\ref{fig:three arenas with different topologies}) for the experiment. Each obstacle is considered as a ``hole'' in the environment -- topological space $X$. The number of ``holes'', as discussed above, is known in topology as the first Betti number $\beta_1(X)$. Our task is to retrieve this number $\beta_1(X)$ from neural cells' activity.

There are two different approaches to this problem which we tested on our experimental data.  The detailed description is provided in an Appendix. 
In short, we propose two approaches and the relate them in a way Downker \cite{Dowker} did it for two given complexes back in 50s. 
The first approach is based on coactivation detection in a given neural group ($n$ chosen neurons) and that allows to construct an element of a simplicial complex of according dimension. This complex might be further studied to retrieve topological invariants from it. 
The second approach is about analyzing the point clouds made of individual neurons' activities. For every point the nearest neighbors are chosen and after that a graph of them is being constructed. This $n$-dimensional graph is then scaled down to a dimension where it might be analyzed without significant information loss. The topological invariants of this graph are also retrieved and analyzed then. 

For more information, see \nameref{S1_Appendix}.

\section{Materials, Experiments and Data Processing}

It takes several preliminary preparations to make the input data suitable for testing our hypotheses. The whole experiment and its processing can be divided into 4 separate stages:

\subsection{Design of the experiment}

4 male C57Bl/6J mice aged 3 months at the start of the experiment were taken for this study. Prior to imaging, animals underwent two successive surgical manipulations: 

-- first, viral vector coding of the GCaMP6s calcium indicator was delivered to the CA1 field of the hippocampus. The animals were anesthetised with a zoletil-xylazine mixture (40 and 5 mg/kg, respectively) and fixed in a stereotaxic holder (Stoelting, Wood Dale, IL, USA). Then a circular 2-mm-diameter craniotomy was made (Bregma: -1.9 mm AP , -1.4 mm ML), and 500 nl of rAAV viral particles (AAV-DJ-CAG-GCaMP6s) were injected at a depth of 1.25 mm from the brain surface. Injections were performed through a 50 $\mu$m tip diameter glass micropipette (Wiretrol I, 5-000-1001, Drummond, USA) by UltraMicroPump with Micro4 Controller (WPI Inc., Sarasota, IL, USA) at a rate of 100 nl/min. After the injection, all exposed surfaces of the brain tissue were sealed with KWIK-SIL silicone adhesive (WPI). 

-- two weeks later, the animals were anesthetized and fixed in stereotaxis again, the silicone cap was removed, and the dura matter was perforated and gently removed from the craniotomy site. Then, a 1.0 mm diameter GRIN lens probe (Inscopix Inc., Palo Alto, CA, USA) was lowered slowly to a depth of 1.1 mm while constantly washing the craniotomy site with sterile cortex buffer. Next, all the exposed brain tissue was sealed with KWIK-SIL, and the lens probe was fixed to the skull surface with dental acrylic (Stoelting). 

-- another two weeks later the animals were checked for a fluorescent calcium signal under light anesthesia (1/2 of the dose described above). The mice were fixed in the stereotaxis and NVista HD miniscope (Inscopix) was lowered upon the GRIN lens probe and the most optimal field of view was chosen. Then a baseplate for further chronic imaging was affixed to the scull surface with dental acrylic.

-- finally, after 1 week recovery, awake mice with attached NVista HD miniscope were put into a custom made circular O-shaped arena (50 cm diameter, 5 cm width, with 5cm high borders) with proximal (different border material) and distal (placed on surrounding curtain 20 cm apart from track) visual cues. 

Each mouse was studying arenas environments for approximately 10-12 minutes while it's neural activity data was recorded on camera through a miniscope at 20 fps. At the same time it's physical activity on the arena was tracked using camera fixed above arena with the same frame rate.

\subsection{Collected data processing and analysis}

The two video streams were first processed.
For calcium imaging data streams it included cropping, stabilizing the image (both laterally and by height), removing pixel noise and adjusting brightness and contrast parameters across the whole frame (polynomial background removal). In case of videos with a moving rodent we also made curve color scheme adjustment to make the background as bright and the mouse as dark as possible. The same preset was used for all the experiments. It all was mainly performed using CaImAn open source Python library~\cite{CaImAn} and KdenLive open source video editor. 
Then, using the same software, the neuron candidate spikes were identified. The method CNMFE-E (Constrained Nonnegative Matrix Factorization for microEndoscopic data) was used for effectiveness. It takes origin from works~\cite{CNMF1,CNMF2} and increases performance significantly as it is looking for neurons in little zones and then merges the whole data framewise.
The data is analyzed by several criteria to filter out the most probable candidates to be actual neurons. The criteria were: 

-- signal intensity (shouldn't be lower than a predefined threshold value);

-- the shape of the spike on each frame (it should be roundish, evenly lightened radially, and monotonously faded by the edge);

-- the shape and the duration of the impulse created by the spike (it should be no longer than a threshold value -- about 2 seconds at most -- and should be initialed rapidly and attenuate monotonously, smoothly and evenly on the edge). The candidates which fluorescent permanently were removed (most likely those are blood vessels);

-- similarly behaving and closely situated candidates are considered as one neuron.

\begin{figure}[!htb]
\centering
\includegraphics[width=0.9\linewidth]{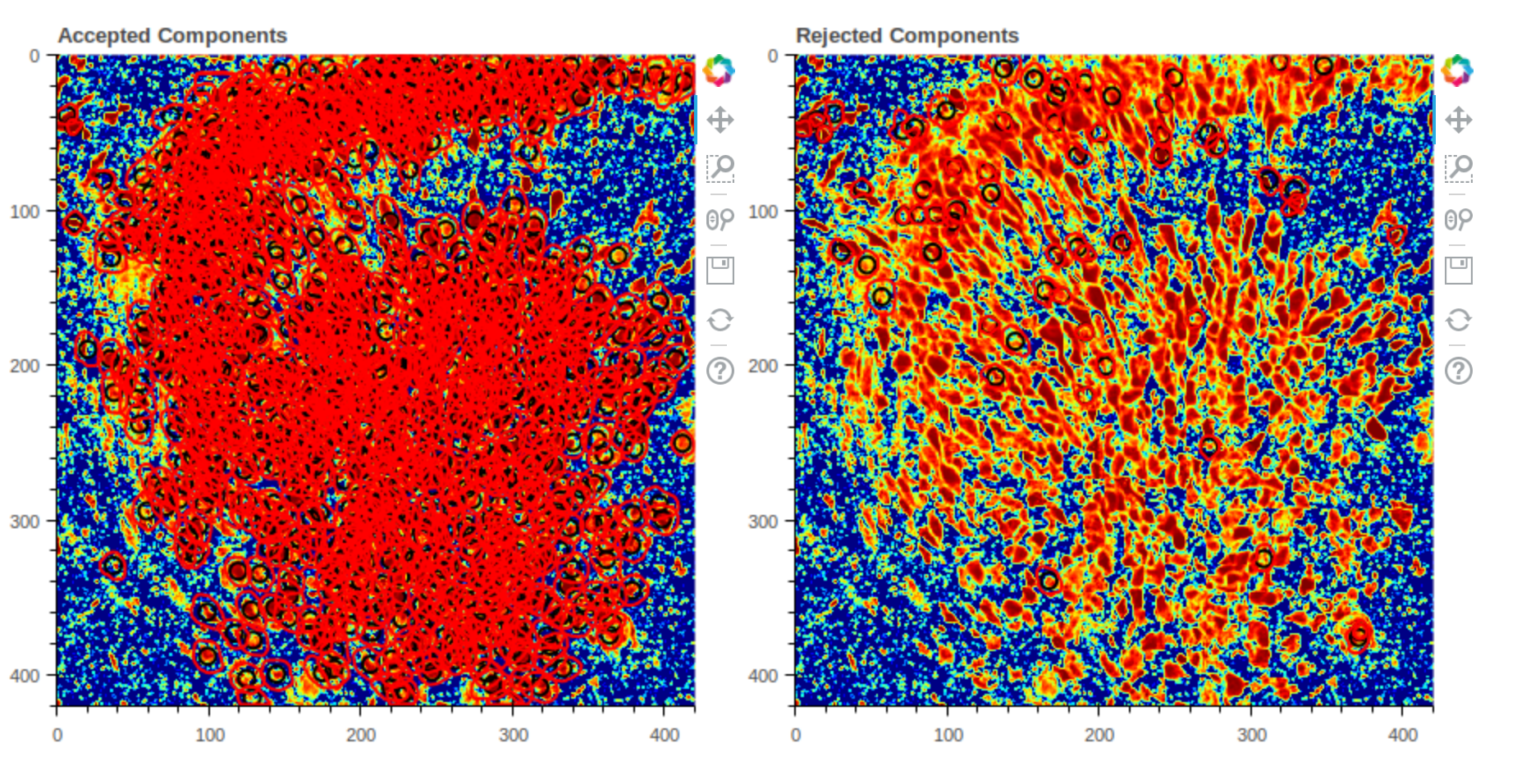}
\caption{Locations and the amount of accepted and rejected candidates to neurons. Red areas have larger density of spike sources. Axes scale unit is 10$\mu m$.} \label{fig: accepted_rejected}
\end{figure}

The ratio of accepted and rejected candidates to neurons is approximately $10:1$ accordingly (numberwise, most of the time it was about 500 overall ``real'' and 50 fake neurons. There were a couple outlier cases with significantly less neurons although the proportion remained consistent). A typical example can be seen on Figure \ref{fig: accepted_rejected}.

\begin{figure}[!htb]
\centering
\includegraphics[width=0.7\linewidth]{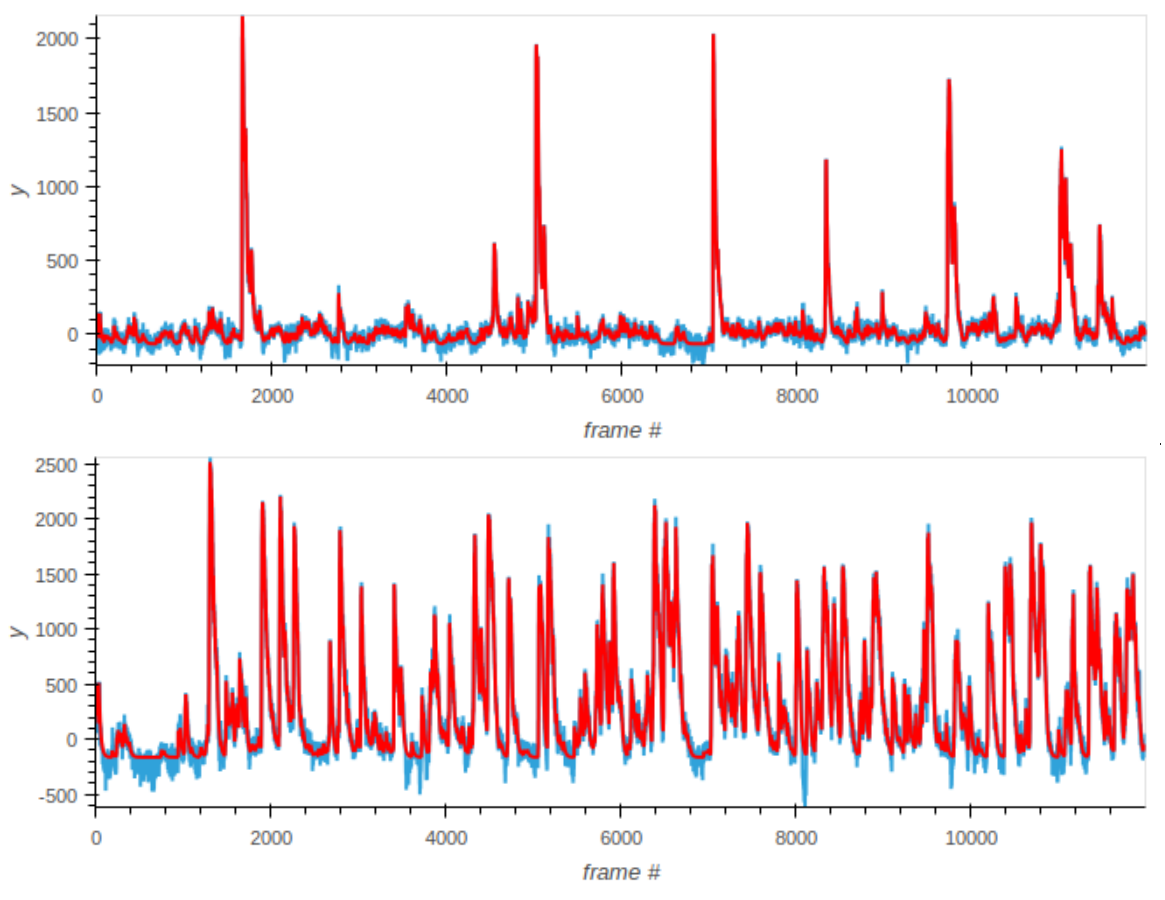}
\caption{Accepted neurons' activities examples. The $x$-axis shows the frame number (frame rate is 20 fps), the $y$-axis shows unnormalized pixel brightness. Red signal is the denoized one, the blue one is the original.} \label{fig: traces_activity}
\end{figure}

Then the activity data of each chosen neuron cell was saved into a table, an exemplar time series can be seen on Figure \ref{fig: traces_activity}.

For the video with the moving mouse the processing included also cropping, stabilizing and applying several effects in KdenLive. This was done to adjust the brightness and the contrast levels to make the animal the darkest and largest object on the arena and make the camera chord ``invisible''. Also, the average empty arena frame was taken, blurred and subtracted from each frame with the moving mouse. Then the largest and darkest object was found on each frame, and its center of mass was calculated. Each center of mass position was saved and joined to form the mouse's trajectory. If some frames were missing or the distance between points exceeded some threshold value (the animal's position between two successive frames changed by more than 25 pixels), then the mean of coordinates between two ``good'' frames were taken. 

\begin{figure}[!htb]
\centering
\includegraphics[width=0.5\linewidth]{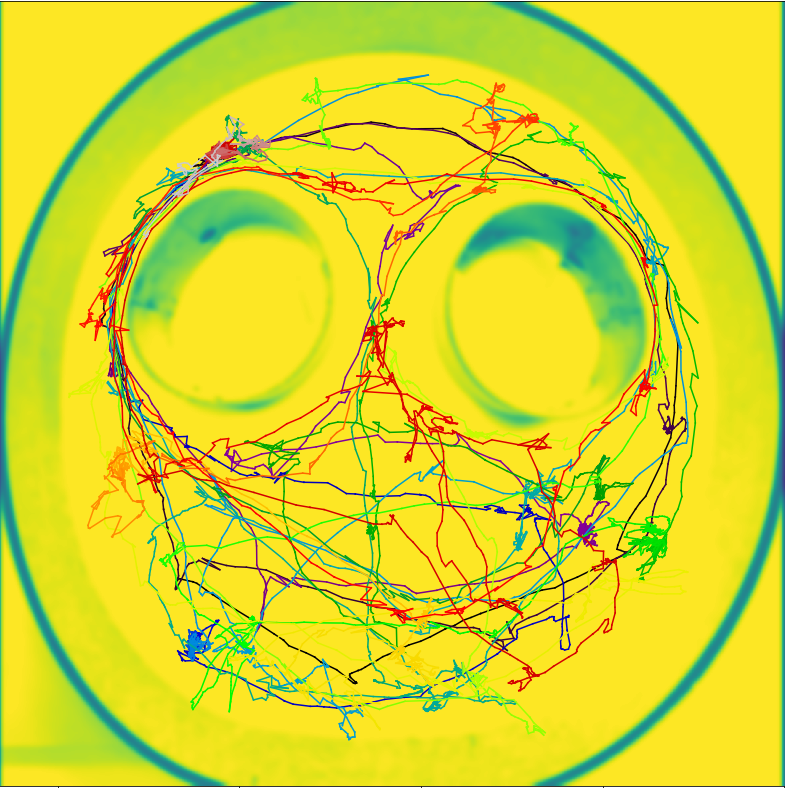}
\caption{An example of the animal's trace during the experiment.} \label{fig: traces_over_arena}
\end{figure}

The obtained trace is now a table, which can be analyzed pointwise, to find out how much time the animal spent in each particular place, to calculate its pace and some other parameters if needed, see Fig.\ref{fig: traces_over_arena}.

Then the data obtained from each video -- activity of chosen place cells and mouse's trace -- were merged and synchronized.

\subsection{Searching for place fields and place cells}

After preliminary processing, we normalize neural data (so that each cell's activity takes values from 0 to 1) and binarize it (the neuron is considered active ($=1$) if its current activity is higher then a certain threshold value 
\begin{center}
$ \text{threshold}  \geq 0.9*(maximal \hspace{0.1 in} level \hspace{0.1 in} of \hspace{0.1 in} brightness)$). 
\end{center}

\rem The $0.9$ factor is, of course, an heuristic value but further experiments have shown that it doesn't affect further data analysis dramatically even if chosen to be as low as $0.4$ because the neurons, being far from binary, were never seen to be half-active. 

\begin{figure}[!htb]
\centering
\includegraphics[width=0.5\linewidth]{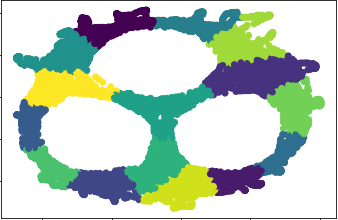}
\caption{A symmetrically binned arena.} \label{fig:track_segment}
\end{figure}

After that we start searching for place cells candidates. As we have two independent random inputs -- the neuron's activity and the animal's position, we tested two commonly used statistical approaches -- based on spatial and mutual information. To discretize the coordinates' input and make it similar to the neural boolean type, in both methods the arena was divided into segments -- so-called bins. \ref{fig:track_segment}. These bins were chosen in a centrally-symmetrical fashion, which might not be a good idea if the arena is not centrally symmetric, but in our case at least one arena -- the one with three obstacles -- is indeed such and the one with two obstacles is close to that.

--- Mutual information ~\cite{mutual} ($\text{MI}$) is a quantity that measures mutual dependence between the animal's discretized position (r.v. $X$ taking values $\hat{x}_i \in \{0,1\}$ with $i \in  [1, \dots, n ]$) and the activity of particular neurons (r.v $Y$ taking values $y_j\in\{0,1\}$ with $\in Y= [1,\dots, m]$). It is calculated in the following way:

\[ \text{MI}(X,Y) = \sum_{i} \sum_{j} P(\hat{x}_i, y_j)  \log \left[ \frac{P( \hat{x}_i,y_i)}{ P_{X}(\hat{x}_i)P_{Y}(y_i)} \right] \]

MI quantifies the dependence between the joint distribution of $\hat{x}_i$ and $y_j$ and what it would be if they were independent. 

--- Spatial information (SI) ~\cite{spatial} can be thought of as the (weighted with $P(\hat{x}_i)$) KL-divergence between 1) $E(\hat{x}_i, y)$ -- the expectation of the indicator (and thus the probability) of the animal being at position $\hat{x}_i$ and the neural signal taking value $y$, and 2) $E(y)$ -- the probability of the neuron being active independent of the animal's position:

$$ \text{SI}(X, Y) = \sum_{i=1}^{n} P(\hat{x}_i) E(\hat{x}_i, y) \log \left[ \frac{E(\hat{x}_i, y)}{E(y)} \right] $$

\begin{figure}[!htb]
\centering
\includegraphics[width=0.9\linewidth]{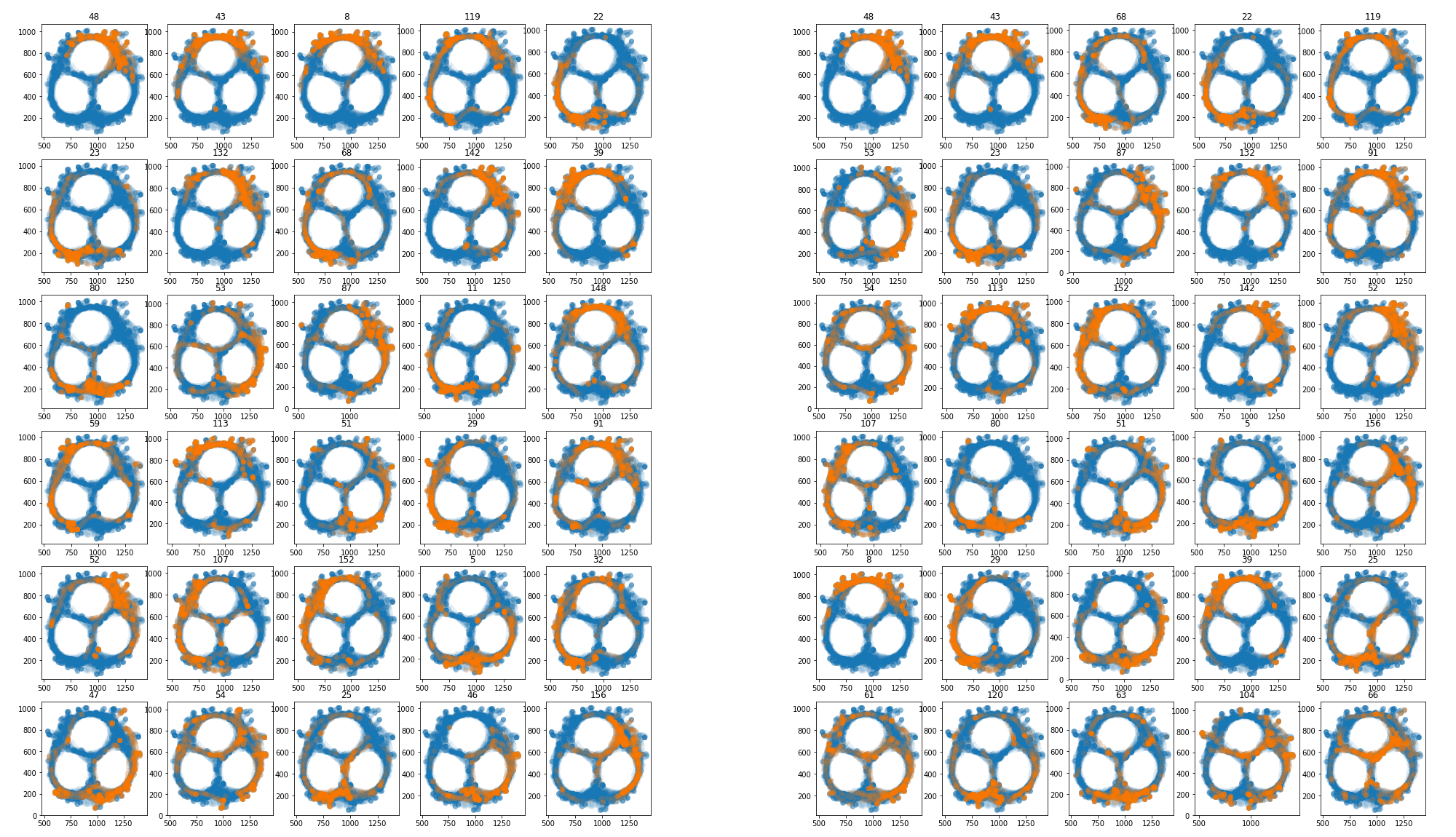}
\caption{Mutual Information (left) vs Spatial Information (right) place cells candidates' PFs comparison.} \label{comparison}
\end{figure}

These methods both might prove useful. We made a series of comparisons (Fig.\ref{comparison}) where we've chosen place cell candidates with both methods and plotted their place fields (in orange) over all the animal's tracked points (in blue).
From this one can clearly see that in our case, SI and MI give similar results.

Nevertheless we did not limit ourselves to one particular method -- but rather used the intersection of active neuron sets provided by both, which filtered out some of the original neurons.

The animal sometimes tends to sit still at some point of the arena -- grooming, sniffing, orienting, etc. That's why place cells could be looked for independently only at moments of time when the animal was moving in the environment. This was reasonable due to the following:

--- By that, we potentially drop ``fake'' place cells related to the other activity;

--- We normalize the distribution of time spent in all the bins and increase the amount of data points where the mouse moved through the center of the arena. Rodents don't like finding themselves in open environments -- they tend to hide near the walls, but the center is crucial for the topological problem that we are solving, and it should have the largest possible weight in the data. 

\begin{figure}[!htb]
\centering
\includegraphics[width=0.9\linewidth]{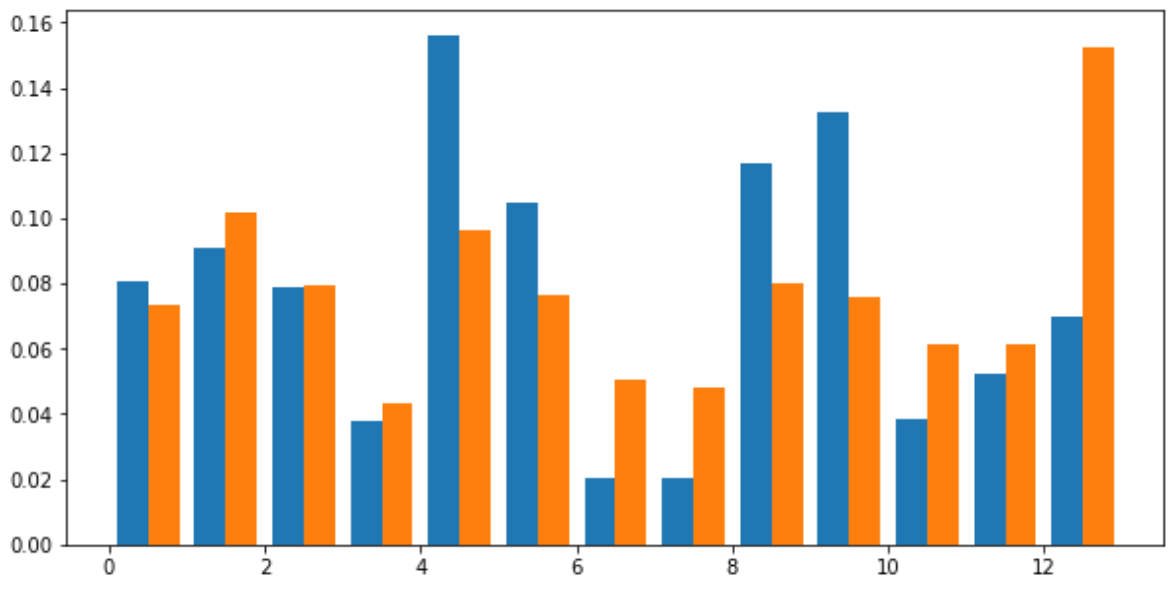}
\caption{Normalized frequencies of video frames spent by the tracked mouse in different bins. All frames (blue) vs frames when the mouse was moving (orange). The rightmost bin is the arena center.} \label{moving_vs_stationary}
\end{figure}

The result of such modifications is presented at Fig.\ref{moving_vs_stationary}. Each bar in the diagram is the normalized amount of time the animal spent if the given bin. The rightmost is the bin related to the center of the arena. As we compare all the moments of tracking (blue bars) to the moments when the mouse was moving (orange) we see that the goal is achieved -- the amount of considered points in the center of the arena (the rightmost bar) increased significantly and the distribution across all the other bins got more even.

To better identify place cells with MI and SI only from the frames when mice were moving, we intersect these with the set of neurons we've chosen at the previous stage. With this done we potentially filter out speed-related cells from the chosen neurons set. All these operations allow to cut off about a half of the initial number of neurons. This approach payed off later when working on dimension reduction.

\section{Results}

\subsection{The Nerve theorem}

In the previous part we've already shown how the PFs of neurons we picked are covering different parts of the arena. Now we can construct a filtration of the simplicial complex that encodes the neurons' mutual activities. Assume one adds an edge to the simplicial complex (SC) if two neurons were both active for longer that a certain time threshold. 

Although the amplitude thresholds (binarizing the activity) do not globally affect the shape of neurons' PFs, they significantly affect the temporal structure of the resulting SC, because peak levels of different registered spikes vary noticeably.  

Also worth noting (it is well-known \cite{Ono}), neurons often play different roles simultaneously, appearing in different functional systems of the cognitome. That's why, if we consider activity levels of an animal's neurons during all the time of the experiment, we'll notice that some neurons in summary are active all over the arena. These won't help much in making a covering as we'll have parasitic edges in a simplex not related to a real neuron's PF. Moreover, we can further reduce the number of neurons considering group activity of several neurons. 

Surely, in order to drop very long-living cycles which would definitely affect our observations, we only consider moments when the mouse was moving. The dataset of chosen neurons is as above. The filtration we've built depends on the amplitude threshold (from 0.99 to 0.6). Also we considered the resulting cycles' lifetimes and their robustness, and took top 10-20 longest-living cycles.

The results of several experiments we constructed the cycles for are presented in the figures below.

\begin{figure}[!htb]
\centering
\includegraphics[width=0.9\linewidth]{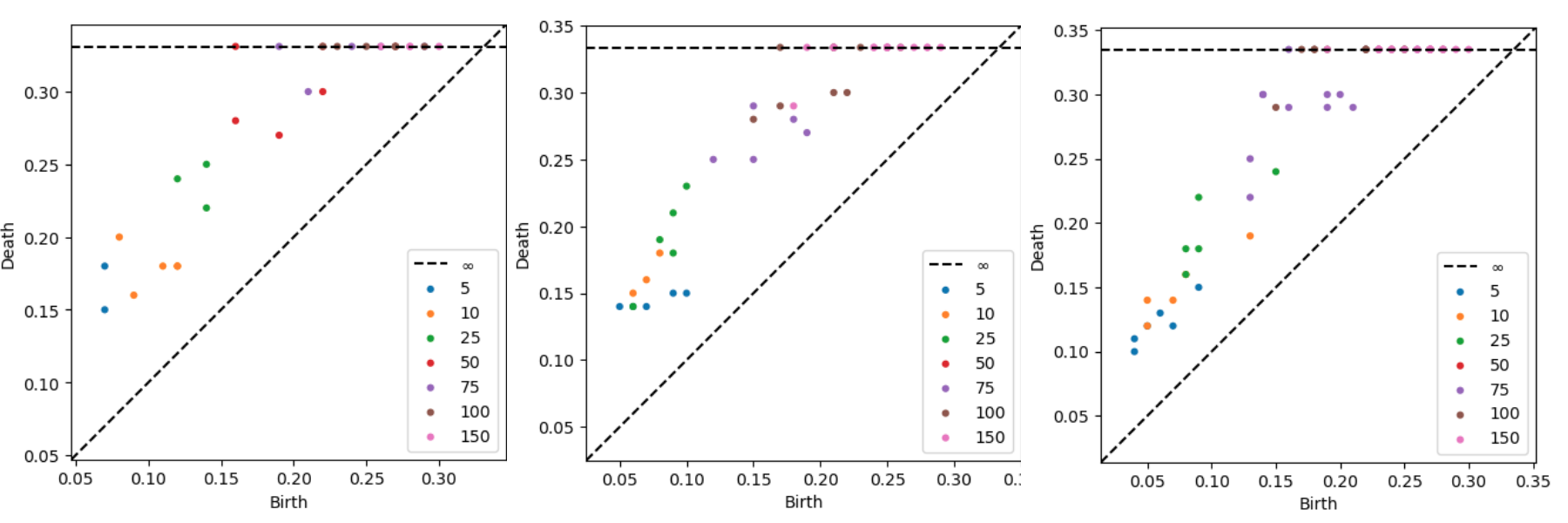}
\caption{PH diagrams of data from neural activity corresponding to three different mice discovering a space with $\beta_1=3$}
\end{figure}

\begin{figure}[!htb]
\centering
\includegraphics[width=0.9\linewidth]{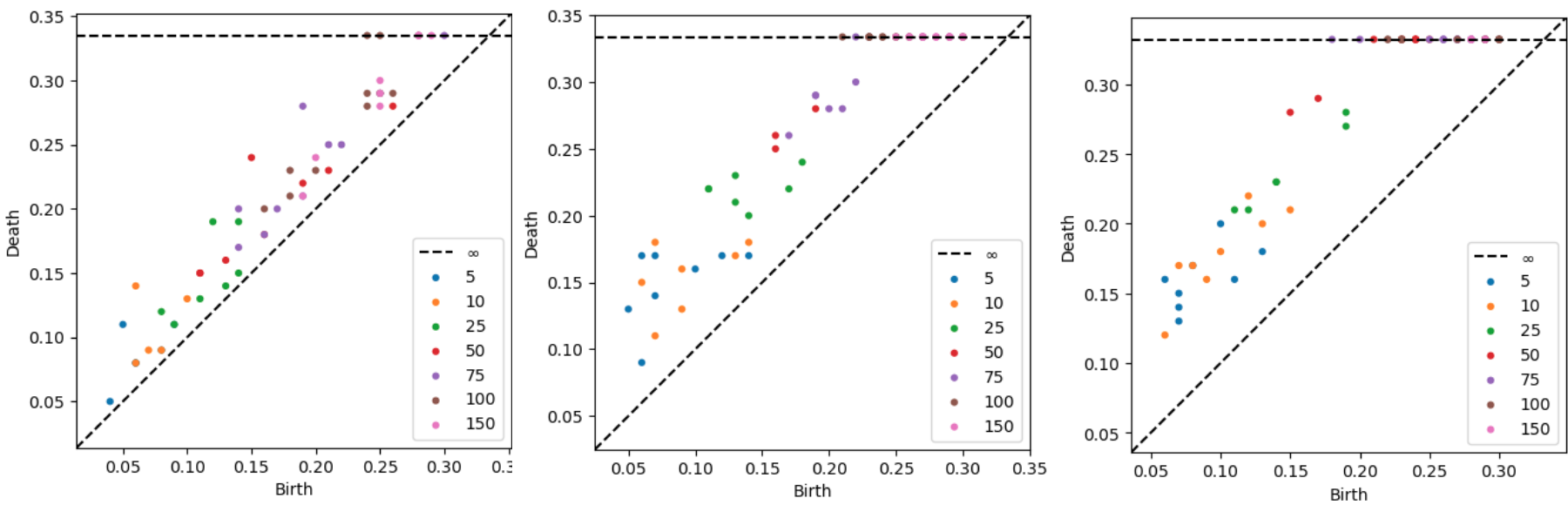}
\caption{PH diagrams of data from neural activity corresponding to three different mice discovering a space with $\beta_1=2$} 
\end{figure}

\begin{figure}[!htb]
\centering
\includegraphics[width=0.9\linewidth]{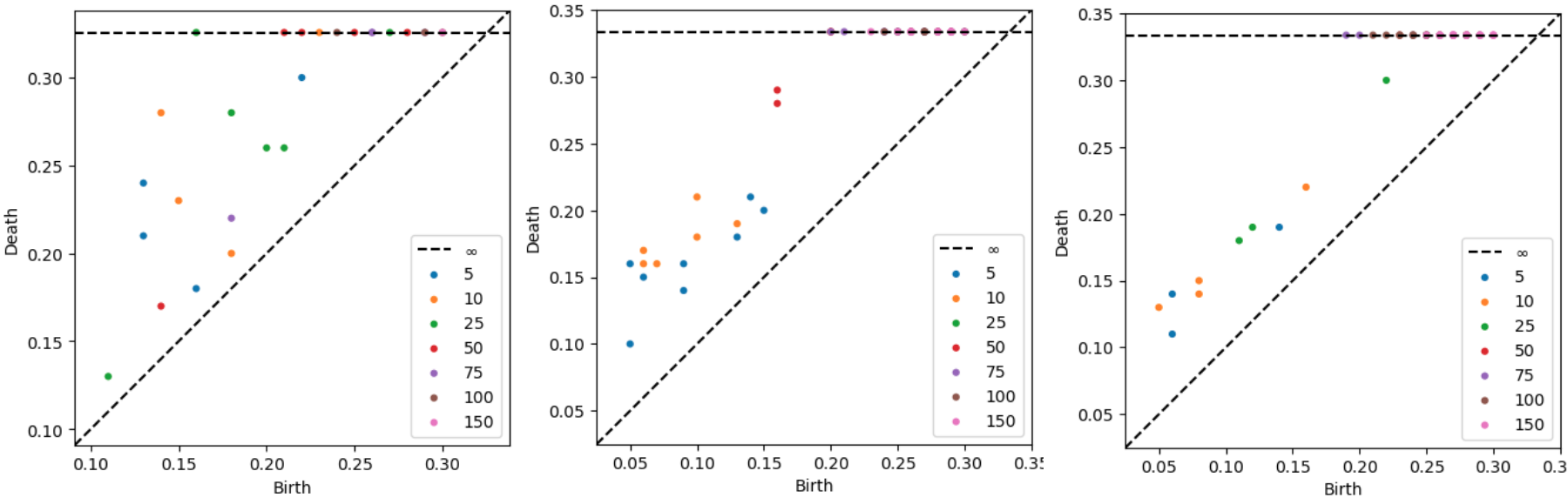}
\caption{PH diagrams of data from neural activity corresponding to three different mice discovering a space with $\beta_1=1$} 
\end{figure}

These images show lots of robust cycles, staying constant for long periods of time. And although there are quite many parasitic ones, compared to the number of holes in the arena, the number of loops the mouse can run along is in fact larger than the Betti number (mouse can use three- or  two-whole loops). Moreover, the cycles on the diagram that are situated close to each other with very similar birth and death times might be united as they are parallel cycles on the graph of states.

\subsection{PCA reconstruction}

We consider the configuration space, where each dimension corresponds to the state of a neuron (a real number from 0 to 1). 

First, we recolor the mouse's trajectory points based on their proximity. 

After we obtain point clouds for each frame we make a projection (PCA) to the 2 dimensional space, the topology of which we are studying. 

\begin{figure}[!htb]
\centering
\includegraphics[width=0.7\linewidth]{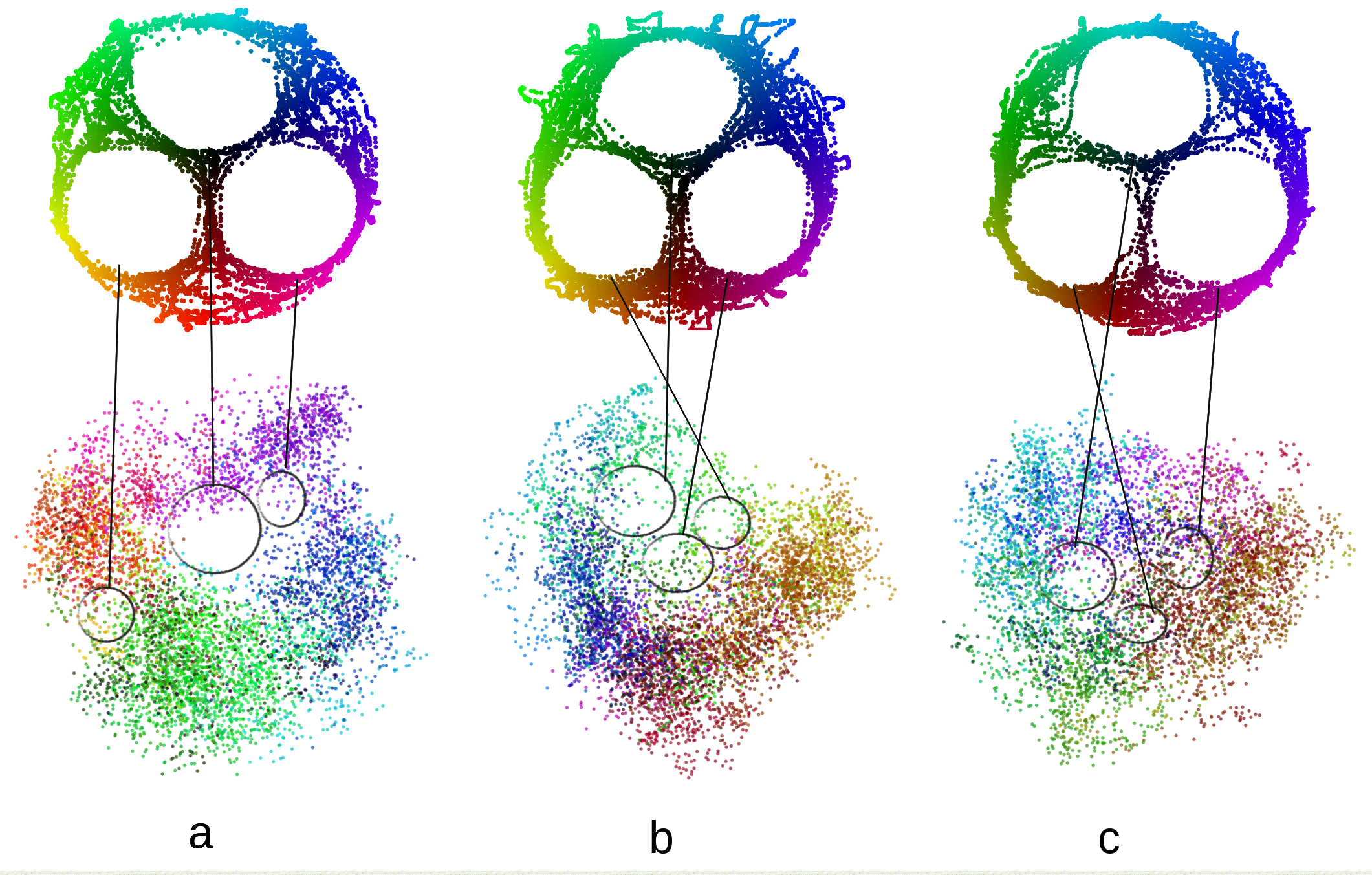}
\caption{PCA reconstruction of the space with $\beta_1 = 3$ from the point cloud of neural activities.}
\label{First Betti number is three for different experiments with different mice} 
\end{figure}

One can clearly see that there are sparse zones where the points are less dense or even vanish completely. This obviously isn't any kind of numerical proof that a hole in the space is reproduced by a group of neurons, but let us give some supportive comments. From Fig.\ref{First Betti number is three for different experiments with different mice} one can see that lesser dense regions of point clouds could be related to points but sometimes the ``holes'' we are looking for are united into one and in general are less pronounced than expected. Colors give further numerical support for the validity of this method. 

\begin{figure}[!htb]
\centering
\includegraphics[width=0.7\linewidth]{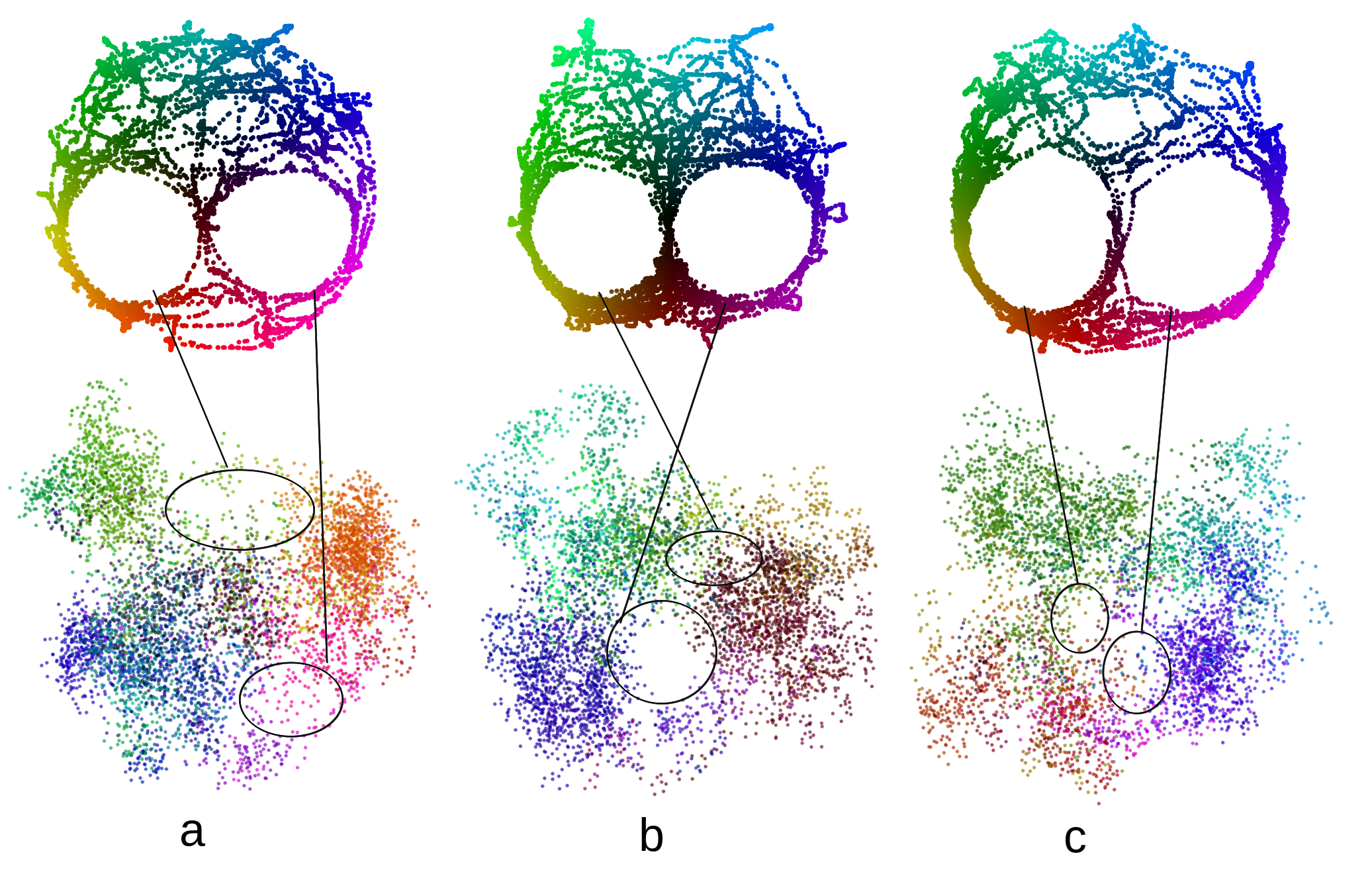}
\caption{PCA reconstruction of the space with $\beta_1 = 2$ from the point clouds of neural activities.}
\label{First Betti number is two for different experiments with different mice} 
\end{figure}

\begin{figure}[!htb]
\centering
\includegraphics[width=0.7\linewidth]{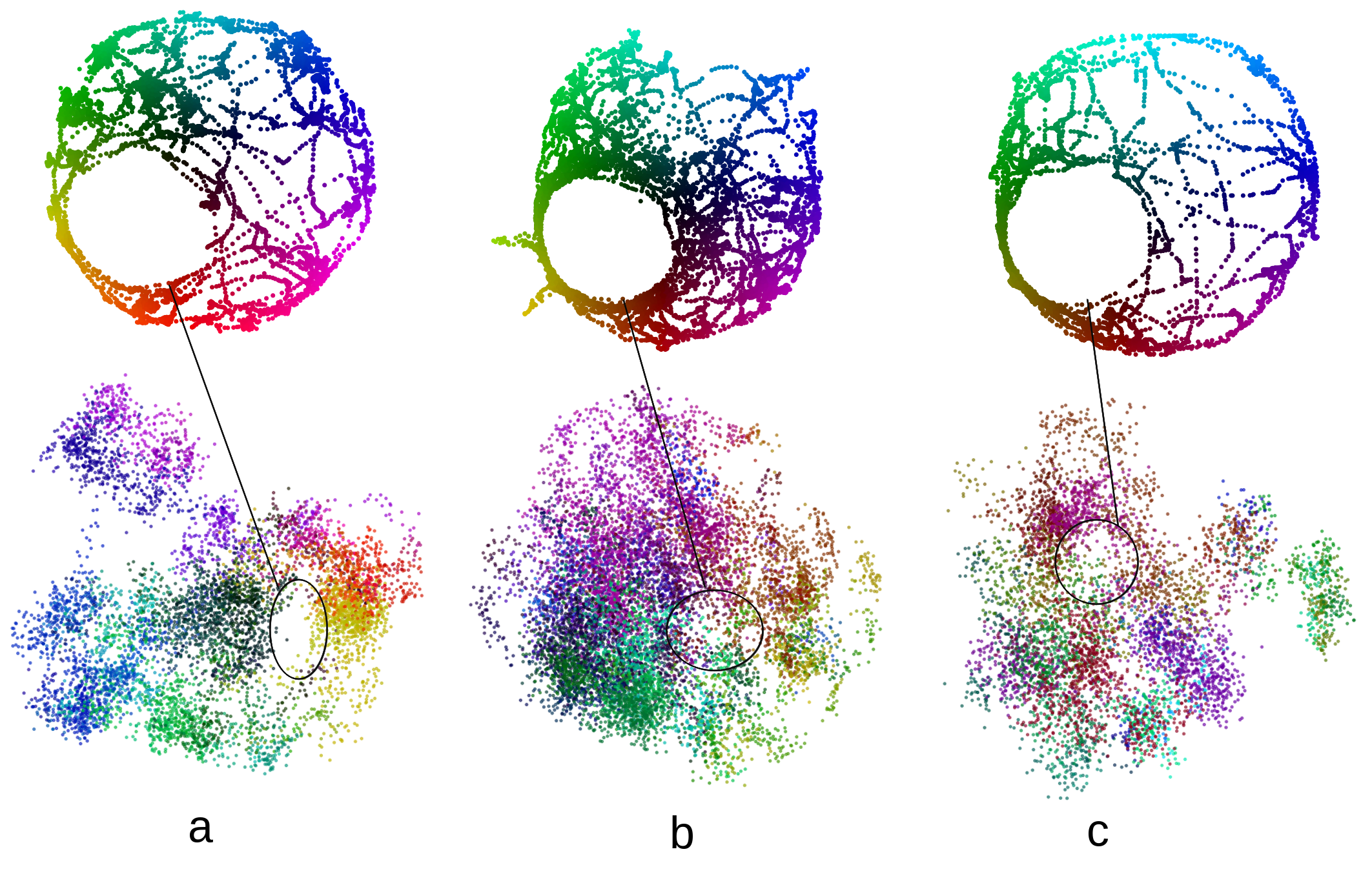}
\caption{PCA reconstruction of the space with $\beta_1 = 1$ from the point cloud of neural activities.}
\label{First Betti number is one for different experiments with different mice} 
\end{figure}

\subsection{Other dimension reduction approaches}

Beyond the obvious PCA methods of the point cloud dimension reduction, one could use more sophisticated ones, like T-SNE, Isomap, Mapper, Locally Linear Embeddings, etc. These are based on different approaches but theoretically (and later it proved to be so practically), Isomap suited us best. It is based on the idea of K Nearest Neighbors at the high dimensional stage and constructs a weighted graph from geodesic distances between data points. Indeed, the data in a brain is unlikely Euclidean. After that the algorithm performs Multidimensional scaling (MDS) of the obtained high-dimensional loops. 

\subsection{Computing persistent homology of embeddings}
\begin{figure}[!htb]
\centering
\includegraphics[width=0.9\linewidth]{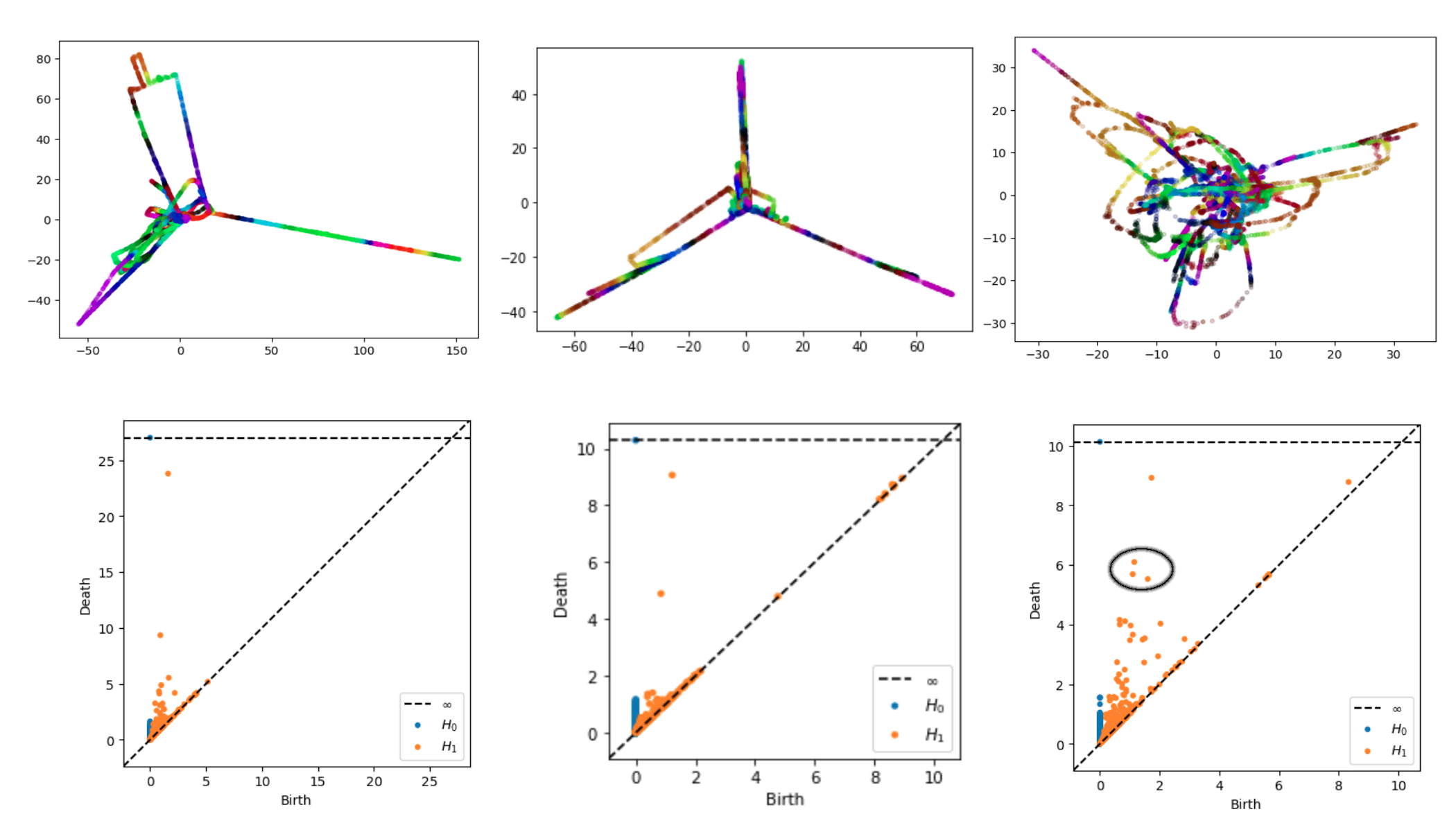}
\caption{Isomap projections and PH diagrams of data from neural activity corresponding to three different mice discovering a space with $\beta_1=3$.}
\label{iso_ht1}
\end{figure}

Below is a demonstration of Isomap reconstruction of cycles (Fig.\ref{iso_ht1}), that is instructive to compare to the results of the Nerve theorem approach. Note that as on the Nerve-theorem diagrams, points that are close on the PH diagram (also called persistence diagram, PD) relate to parallel loops and so might be joined into one. 

\begin{figure}[!htb]
\centering
\includegraphics[width=0.9\linewidth]{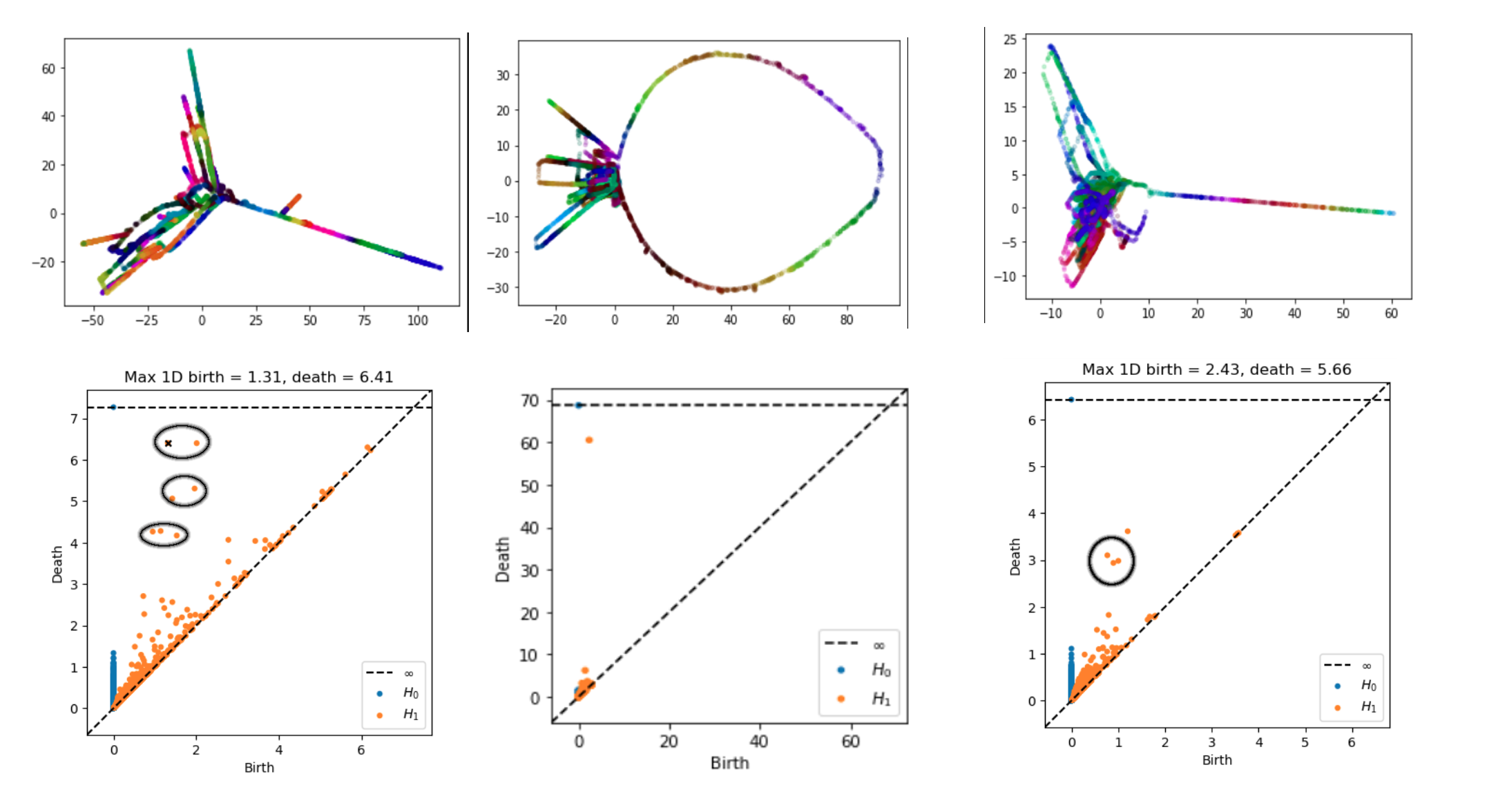}
\caption{Isomap projections and PH diagrams of data from neural activity corresponding to three different mice discovering a space with $\beta_1=2$.}
\end{figure}

\begin{figure}[!htb]
\centering
\includegraphics[width=0.5\linewidth]{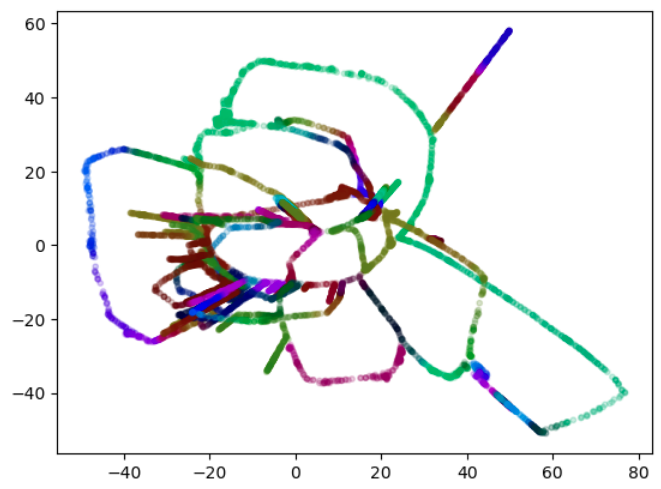}
\caption{Example of Isomap projections and PH diagrams of data from random neural activity.}
\label{iso_shit}
\end{figure}

One can clearly see that Isomap indeed preserves some of the cycles, which are far from being random. Our studies showed that once we choose the neurons from other neural groups, Isomap finds a very high amount of loops (see Fig.\ref{iso_shit}), the reason possibly being that it finds lots of neurons' parallel activities not related to the animal's actual motion. However, it is not a robust method for reconstructing precise topological features from a very limited amount of neurons.

\begin{figure}[!htb]
\centering
\includegraphics[width=0.9\linewidth]{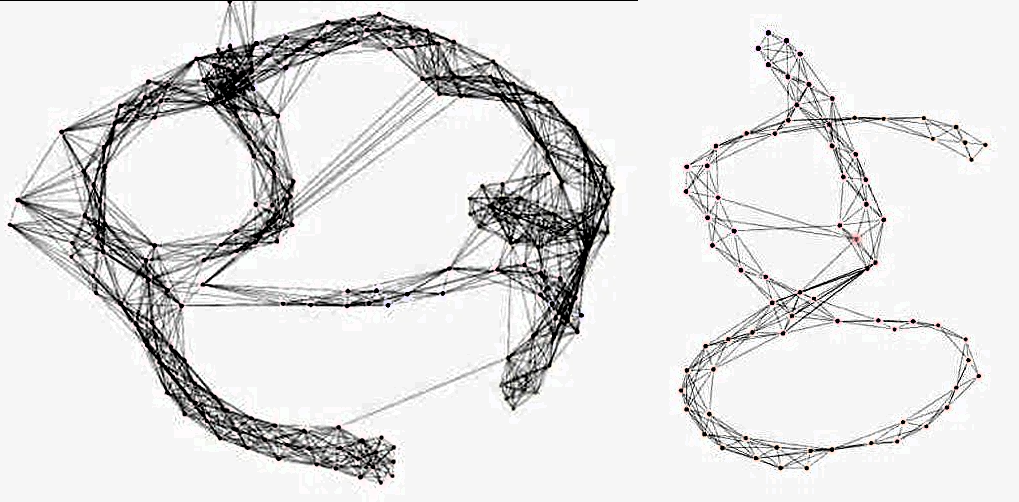}
\caption{Two examples of Mapper output.}
\label{KeplerMapper}
\end{figure}

Among numerous tools for dimension reduction that we tested, only KeplerMapper (with quantile transformer as a scaler and k-Means as a clusterer) gave somewhat similar results with imprecise number of pronounced loops which are however different from what was expected (Fig.\ref{KeplerMapper}).

The above results lead to the idea that the amount o neural data we have is not enough to fully infer the topological properties of the physical space the animal is placed in, even despite the fact that functional systems of neurons somewhat duplicate each other's functions. However, from our data we obtain reproducible results in that the number of loops and cycles in the neural activity is about as small as that in the physical space, which leaves us positive about present research.

\subsection{Diving into the projections}

Projections of nearest neighbors graphs are critical to be understood. In a sense, their weaknesses are as well their strong sides. Speaking of the point clouds, it's obvious that similar combinations of neural activities might appear at different moments of an experiment, which might affect the temporal structure of the neighbors graph - two points, which are close to each other might come from different time moments. 
That's why we decided to explore our data further and, in particular, the behavior of the resulting projections.  

\begin{figure}[!htb]
\centering
\includegraphics[width=0.4\linewidth]{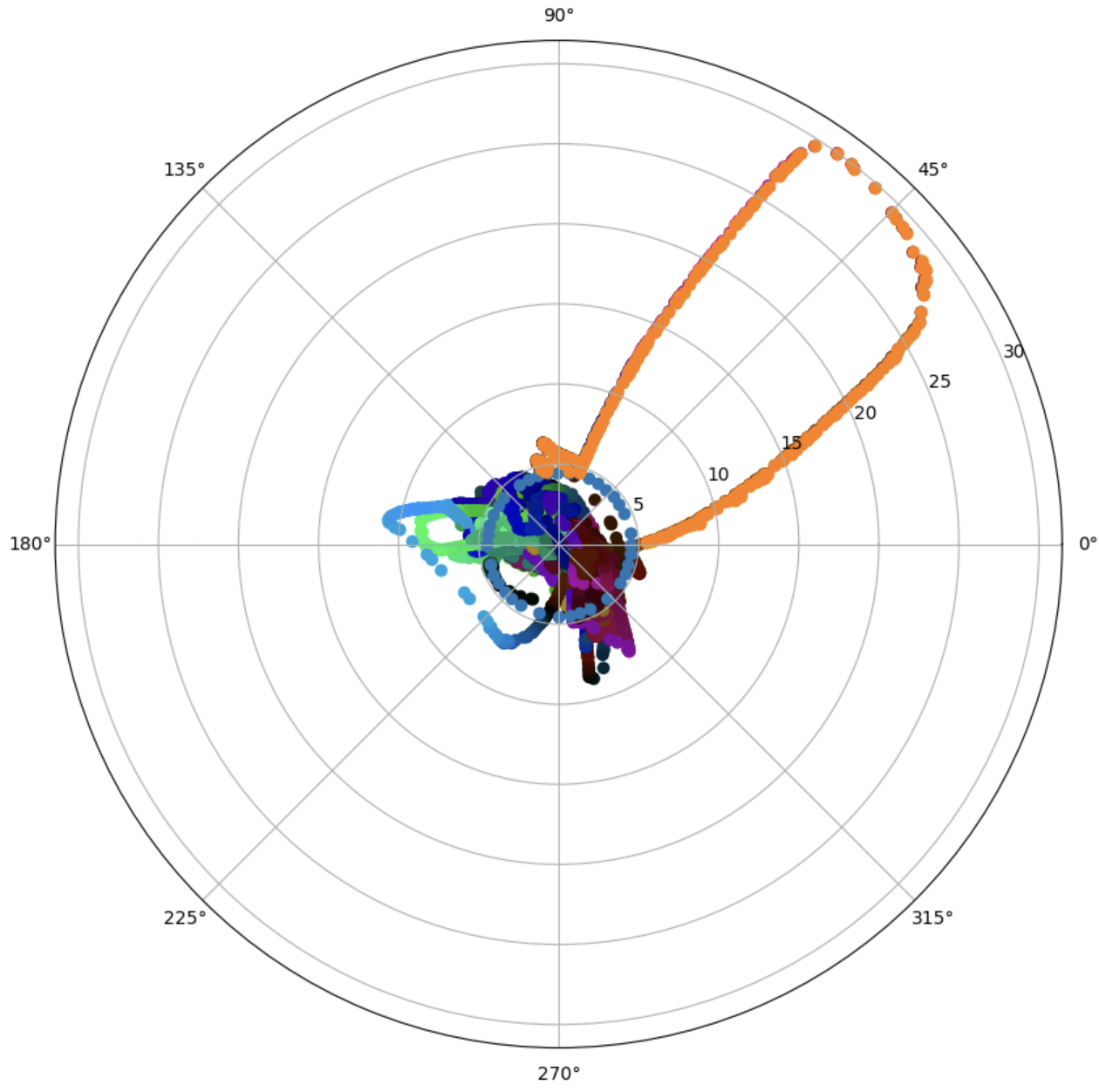}
\caption{One loop extracted from Isomap projection.}
\label{isomap_loop} 
\end{figure}

\begin{figure}[!htb]
\centering
\includegraphics[width=0.5\linewidth]{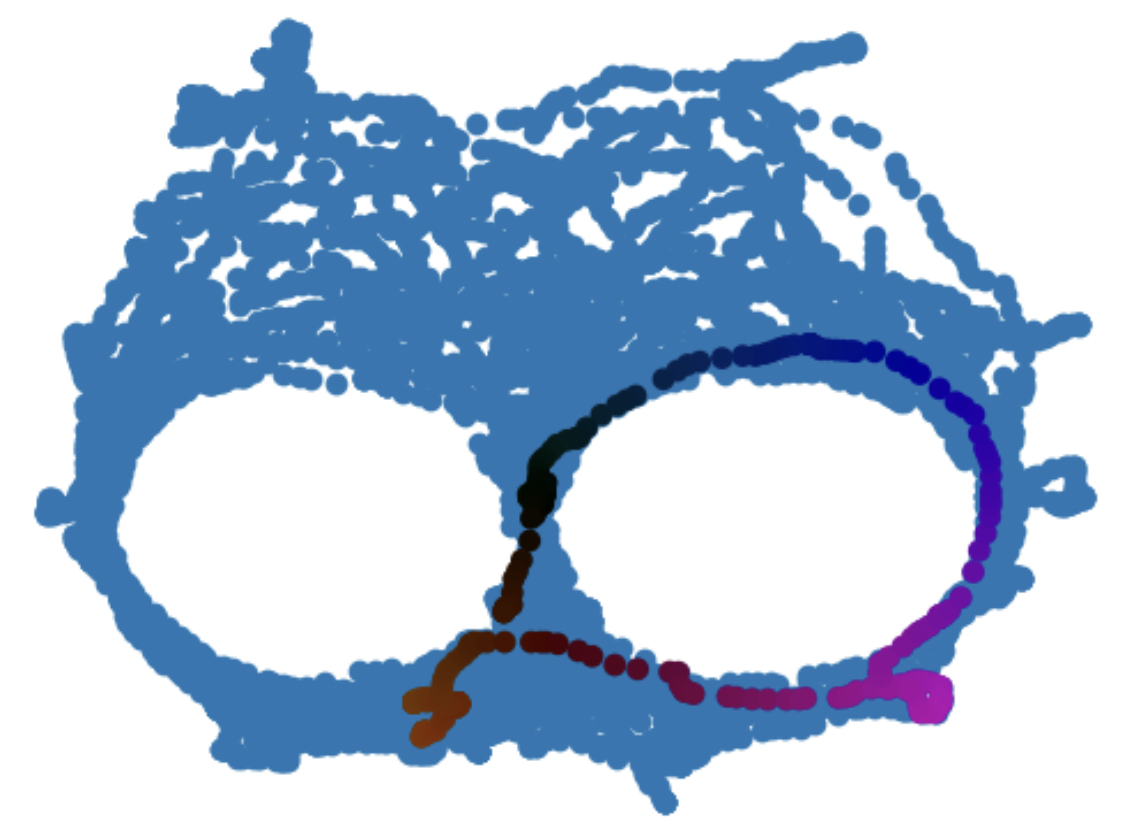}
\caption{Isomap loop points tracked on real mouse trajectory}
\label{isomap_loop_back} 
\end{figure}

One can clearly see that the loop on the nearest neighbour projection corresponds to a loop around the hole on the arena where the mouse navigated. It appeared to be very pronounced, but to approve the usefulness of the nearest neighbours graph projections we decided to make the following study. 

First we found the mass centers of the loops on the arena and then tracked all the activities within some neighborhood around it.  After that we specifically tracked mouse's traces in order to find specific closed loops when an animal physically travelled around the hole at arena. 

After that we take all the possible activities across all the neurons (not only the place cells preliminary chosen above) and make Isomap projections of short time periods when the mouse navigated around an arena hole.  For each such moment we extracted 5-10 active neurons (out of 300-500 overall - the rest were inactive) and built the graph of nearest neighbours for them. Surprisingly, the neurons groups for which Isomap projections gave the blueprint of the mouse's trajectory were almost always the same as the ones chosen with the method described is the sections above. 

\begin{figure}[!htb]
\centering
\includegraphics[width=0.7\linewidth]{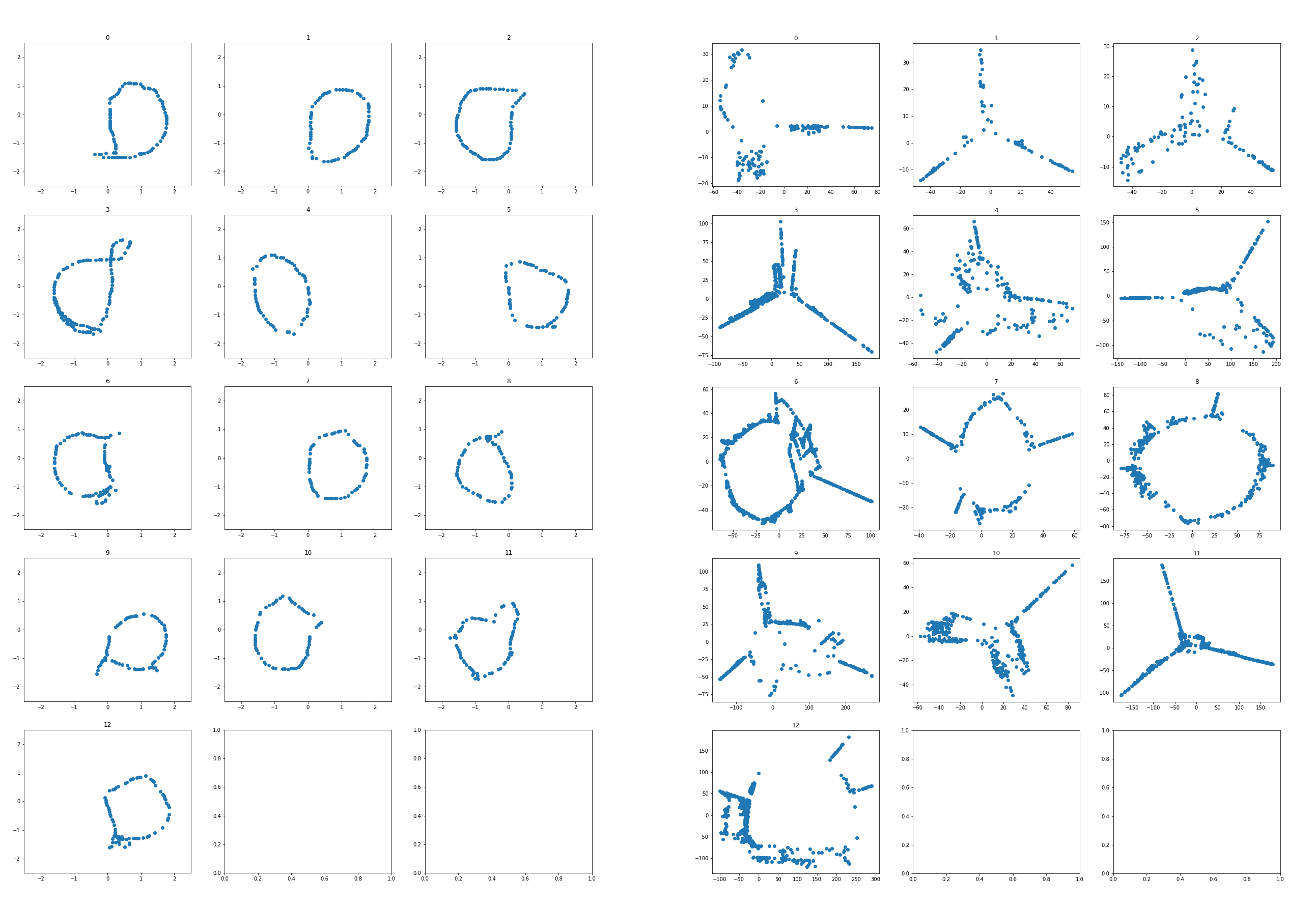}
\caption{Comparison of mouse's original traces around the holes on the arena and the Isomap projections of neural activities at corresponding time moments}
\label{isomap_loops_for_loops} 
\end{figure}

The intersection of active neurons and the ones which were chosen to be place cells is 70\% which is very significant for methods so different. 

\begin{figure}[!htb]
\centering
\includegraphics[width=0.5\linewidth]{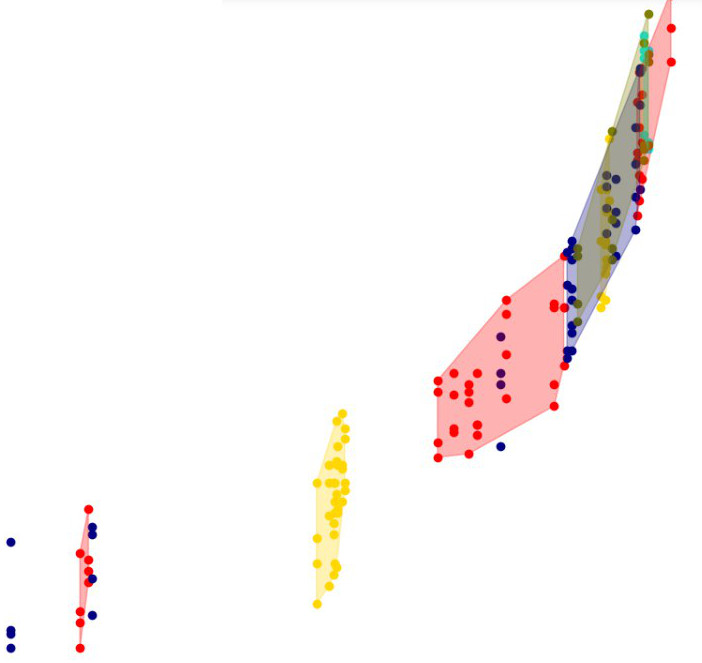}
\caption{Comparison of clusters of neural groups corresponding to top 5 loops chosen by Isomap projections. The closer points are in this plane -- the more similar is their neural activity as a group. Each group of neurons encodes one loop.}
\label{iconvex_hulls} 
\end{figure}

A critical thing to note is the fact that for different time periods one can see different neural groups involved. At the beginning of the experiment and at its ending we see almost unrelated groups at work. 
We made a diagram of neural groups for different loops at different time periods and one can see the intersections of the convex hulls of points corresponding to elements in neural groups are gradually changing \ref{iconvex_hulls}. Each group is denoted by its own color. Still, it's a theme for future detailed study.

\subsection{Combining the approaches (Dowker complex reconstruction)}

Now we can construct a so-called Dowker \cite{Dowker} complex from two previously obtained ones.  The procedure was described in the Appendix below.  Vaguely, the idea is to detect whether there are temporal intersections in the two methods of cycles retrieval.  

Let's choose two of the highest amplitude loops from all the cycles (whose homology groups are of order 1 at most) we obtained when building the time series complex and from higher dimensional nearest neighbors graphs embeddings. 

\begin{figure}[!htb]
\centering
\includegraphics[width=0.7\linewidth]{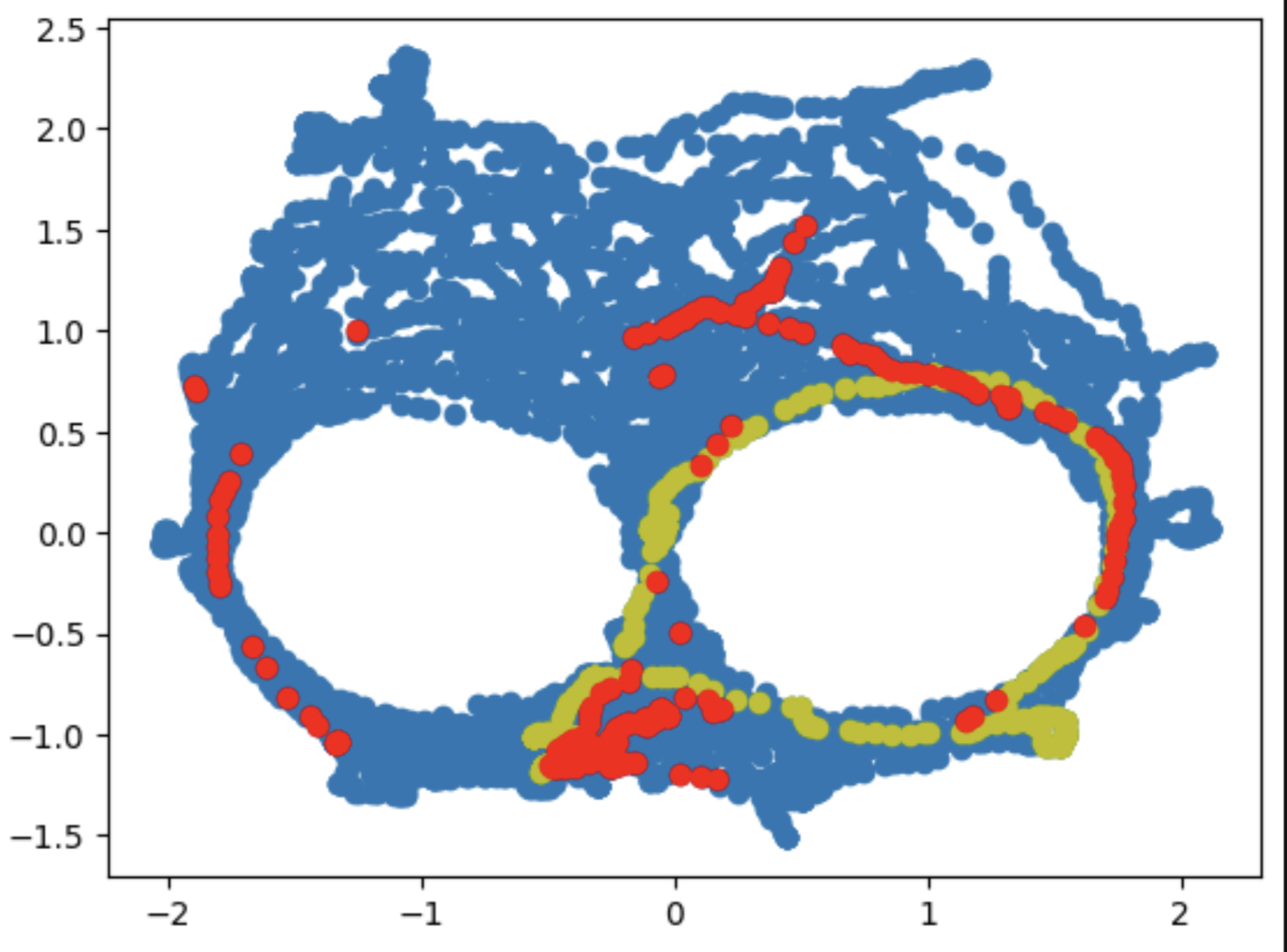}
\caption{Two reconstructions of the loop in the arena space obtained from two complexes retrieved from neural activity data (red is time-series one, yellow is from point cloud analysis) for a particular experiment}
\label{dowk} 
\end{figure}

To illustrate the results of comparison we plotted the points forming the two loops simultaneously, without reducing it to the intersection as it would reduce the number of points to an unnecessarily sparse picture.  What's important here is that we indeed find same loops with two methods -- moreover, two methods complement each others missed points if we choose to use them at the same time. The intersection, however, would still give a pronounced loop in the trace around the  ''hole''  in the arena Fig \ref{dowk}.

\section{Discussion}

Comparing all the results we can notice, especially for the complex topology, like the 3-''hole'' arena, that the most important information the mice's hippocampus captures is if there are any obstacles. 

\begin{figure}[!htb]
\centering
\includegraphics[width=0.7\linewidth]{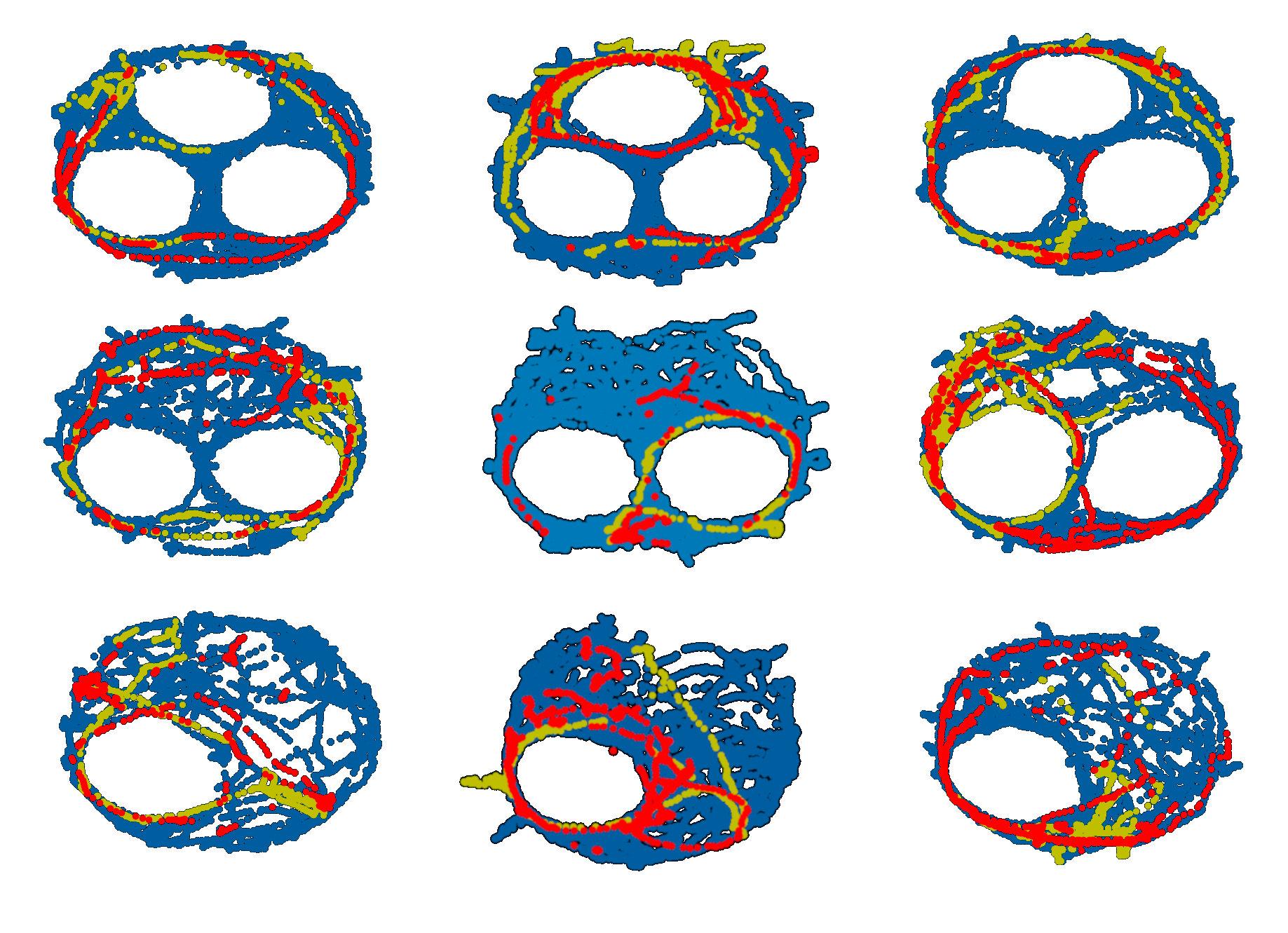}
\caption{Two reconstructions of loops in the arena space obtained from two complexes retrieved from neural activity data (red is the time-series based, yellow is from point cloud analysis) for all the provided experiments}
\label{dowkers} 
\end{figure}

That is in some sense rational, and shows that a mouse might have some issues remembering complex topology but it realizes at least that the space is somewhat topologically non-trivial. \ref{dowkers}. Indeed, it's natural that mice tend to stay closer to the ''walls'' as they feel more ''safe'' this way exploring a new environment.

What's even more interesting is that Isomap projections often have the largest cycle in their graph of nearest neighbors of cells activations corresponding to the trace along the border as well as in time-series complex case. 

\section*{Conclusion}

Despite very optimistic theoretical assumptions, real neural activity data is full of noise, especially considering that we are observing only a limited amount of neurons (just over 500 at best) and it might not be even enough to make a dataset proving the described theory. Sure, we observe some cycles of finite lifetimes, we can construct some close-to-expected homologies either from filtrations of nerve-like complexes or from point cloud data. But we can't build a concrete algorithm that would be universally applicable to any such data and we can't provide a precise algebraic map between two different complexes (like we described in the Dowker complex section). 
We can apply one of our methods or combine them on different datasets and expect the same reproducible results from case to case. We showed that in Dowker complex reconstruction in the last section. The result is that we can find the same patterns in two differently constructed topological interpretations of the result. We can see that the two approaches capture the same cycles related to a real loop in the arena space.  It's obvious that the most successful reconstructions are related to the loops along the walls and they are captured by both our methods which is shown in the Dowker complex approach. 
We intentionally kept the noisy results to show that the task is very hard to reproduce automatically and by now it should not be expected to be solved flawlessly on a small and noisy dataset. 

\section{Supporting information}

\paragraph*{S1 Appendix.}
\label{S1_Appendix}

\subsection{Binary convex code}

The first method was first proposed in the work of Cohn--Dabaghian ~\cite{Cohn}; the theoretical solution is based on the topological Nerve theorem of P.\,S.\,Alexandrov \cite{AlexNerve}. 

In this model the activity of each neuron is treated as binary, which means that at each moment the neuron is either active or inactive (which is not that far from reality (see Section 3)). It is assumed that the activity of a neuron depends only on the animal's position in the physical space $X$, not on the time. This means that a neuron responds in a similar way each time the mouse visits a certain point in space. Let $[m]=\{1,\ldots,m\}$ be the enumerated set of neurons. For each $i\in[m]$ the subset $U_i\subset X$ is specified in which the $i$-th neuron (presumably a place cell) is active. The subset $U_i$ of the physical space is called the place field (PF) of a neuron $i$. It is also assumed that $X=\bigcup_{i\in[m]}U_i$, that is, at each point of $X$ at least one neuron is active. A collection $\ca{U}=\{U_i\}$ of subsets such that $X=\bigcup_{i\in[m]}U_i$ is called a covering of the space $X$. With any covering, one can associate two objects:
\begin{enumerate}
\item The nerve of the covering $\ca{U}$ is the simplicial complex $K_\ca{U}$
\begin{equation}\label{eqNerveDef}
	K_{\ca{U}}=\{I\subseteq [m]\mid \bigcap\nolimits_{i\in I}U_i\neq\varnothing\}
\end{equation}
the simplices of which correspond to collections of subsets $U_i \subset X$ in the following way. With each subset $U_i$ we associate the vertex $v_i$; whenever two subsets $U_{i_1}$ and $U_{i_2}$ intersect, we draw an edge $\{v_{i_1},v_{i_2}\}$; whenever three subsets intersect, we draw a triangle, and so on. 

\item The neural code
\begin{equation}\label{eqNeuralCodeDef}
	\ca{N}_{\ca{U}}=\{\epsilon_x=(x\in U_1, x\in U_2,\ldots,x\in U_n)\in\{0,1\}^n\mid x\in X\}.
\end{equation}
For each point $x$ one forms a binary string $\epsilon_x$ by checking whether $x$ belongs to $U_i$ for all $i\in[m]$. Each binary string $\epsilon_x=(\epsilon_1,\ldots,\epsilon_m)$ corresponds to the subset of indices $A_x=\{i\in[m]\mid \epsilon_i=1\}$. The neural code $\ca{N}_{\ca{U}}$ is, therefore, a collection of subsets $A_x$ of $[m]$ for all possible points $x \in X$. The subset $A\subseteq[m]$ lies in $\ca{N}_{\ca{U}}$ whenever there is a point $x$ of the physical space, where the neurons from the set $A$ are active, while the neurons from the complement $[m]\setminus A$ are inactive.
\end{enumerate}

Both the nerve and the neural code can be obtained from place cells activity, under the assumption that the animal visited all points of $X$. 
The nerve $K_{\ca{U}}$ can be reconstructed from $\ca{N}_{\ca{U}}$:
\begin{equation}\label{eqNerveFromCode}
K_{\ca{U}}=\{J\mid \exists I\in \ca{N}_{\ca{U}}\mbox{ such that } J\subset I\}
\end{equation}
In this sense, the neural code contains more information about neural activity than the nerve. The nerve contains information on which collections of neurons fire simultaneously. The neural code also contains the information about inactivity patterns. 

Recall that there is a notion of homotopy equivalence of topological spaces~\cite{Hatcher}. If two spaces are homotopy equivalent, $X\simeq Y$, their rough topological features, such as Betti numbers, coincide. The space is called contractible if it is homotopy equivalent to a single point. The covering $\ca{U}$ is called contractible, if any nonempty intersection of its elements $U_{i_1}\cap\cdots\cap U_{i_s}\neq\varnothing$ is contractible. The following fundamental result dates back to the work of P.\,S.\,Alexandrov~\cite{AlexNerve}.

\begin{thm}[Nerve Theorem]
If the covering $\ca{U}$ of $X$ is contractible, then the nerve $K_{\ca{U}}$ is homotopy equivalent to $X$.
\end{thm}

Notice that any convex subset $A\subset\Ro^d$ of the Euclidean space is contractible. Therefore, if every set $U_i$ of the covering $\ca{U}$ is convex, then all their intersections are convex as well, hence such covering is contractible. The last assumption of the nerve of place cell activity model states that each place field $U_i$ is convex. Then the Nerve theorem implies the following statement.

\begin{prop}
Assuming that neurons activity is binary, and the place fields $U_i\subset \Ro^2$ are convex, the homotopy type of the physical space $X$ can be obtained from the neural code, since $X\simeq \ca{N}_{\ca{U}}$.
\end{prop}


This approach was significantly extended by Curto, Itskov, et.al. in ~\cite{Itskov} where they proposed the theory of convex neural codes. Recall that the place field (PF) of $i$-th neuron (in the set $[m]$ of neurons) is a subset of the whole topological space $U_i\subset X$ where this particular neuron is active. The whole approach of Curto and Itskov is based on the assumption that all $U_i$'s are convex. The convexity assumption can be weakened though: the Nerve theorem still applies if all place fields $U_i$ as well as their nonempty intersections are at least contractible. 


It happens so though, that the place fields of certain neurons can be either non-connected as well as non-simply-connected, so the Nerve theorem does not apply even theoretically. This difficulty might be resolved in the following way.

One can choose only those cells which have contractible place fields, and apply the algorithm to this collection of neurons. Or instead of using the Nerve theorem, one can apply more general notion of homotopy colimit, resp. the Mayer--Vietoris spectral sequence. These notions allow to reconstruct the homotopy (and  homology) type of the environment $X$, if one knows the homotopy (resp. homology) types of all $U_i$'s, their possible intersections and the types of all possible inclusions of intersections. While theoretically this approach works fairly well, it does not make much sense in the study of cognitive maps, since we do not know how to extract the homotopy type of the place field from the brain activity only. In other words, it is impossible to distinguish intrinsically if the neuron has contractible place field or not. 

\subsection{Point clouds and Persistence}
The second approach one might consider is to look at the neural activity measurements at each time moment as a point cloud in a high-dimensional Euclidean space. This approach was applied in \cite{RBD} to reconstruct the head direction of a freely moving rodent. In that experiment the configuration space is replaced by the circle $S^1$. It was shown possible to extract this circle from the neural data by a combination of principal component analysis (PCA) and topological data analysis (TDA) methods.




Compared to the Nerve theorem approach, we use the ``raw'' (not binarized) neural signals, as actual place cells' activity is far from being binary. After normalization, recorded data is represented by a table $(a_{t,i})$ where each entry is a real number between $0$ and $1$. If it needs to be binarized, one might choose a threshold parameter $\theta\in [0,1]$, and turn each entry $a_{t,i}$ into $0$ if $a_{t,i}\leqslant1-\theta$ and $1$ if $a_{t,i}>1-\theta$. This gives a boolean table, from which the simplicial complex $K_\theta$ can be constructed. This complex, however, depends on the threshold parameter value $\theta$. It can be seen that whenever $\theta_1<\theta_2$ we have an inclusion $K_{\theta_1}\subseteq K_{\theta_2}$. For $\theta=0$, the complex $K_0$ consists of $m$ disjoint points, while $K_1$ is the whole simplex on the set $[m]$. When the threshold parameter $\theta$ changes from $0$ to $1$, the topology of $K_\theta$ changes as well. The theory of persistent homology (PH) is applied when one needs to perform a quantitative analysis of changes in the topology. 

\begin{con}
The collection $\{K_t\}_{t\geqslant 0}$ of simplicial complexes indexed by nonnegative real numbers is called an (increasing) filtration if $t_1<t_2$ implies $K_{t_1}\subseteq K_{t_2}$. It is assumed all the complexes $K_t$ have the same vertex set $[m]$. Since simplicial complexes are discrete objects, only a finite number of changes can occur in a filtration $\{K_t\}$. It is usually assumed that $K_0$ is the set of $m$ disjoint vertices, and the filtration is encoded by a sequence of simplices $\{I_j\}$, added consequently to the complex; and the sequence of their birth-times. The birth-time $t_j$ is the moment when the simplex $I_j$ first appears in the filtration. We note that addition of a $d$-dimensional simplex $I_j$ either creates a $d$-dimensional homology or kills a $(d-1)$-dimensional homology. Therefore, one can track what happens to the Betti numbers at each moment of time.

It is instructive to look at the $d=1$ case: when one adds an edge $I$ to a graph, then $I$ either connects two different components (in which case the number of connected components $\beta_0$ decreases by 1), or creates an additional cycle (in which case $\beta_1$ increases by one). This effect is completely similar in higher dimensions.

The persistent homology of the filtration $\{K_t\}$ is the unordered collection of pairs $\{(\tb_\alpha,\td_\alpha)\}_\alpha$ labeled by nonnegative integers $d_\alpha$. Here $0\leqslant\tb_\alpha\leqslant \td_\alpha\leqslant+\infty$. Each pair $(\tb_\alpha,\td_\alpha)$ reflects the fact that a certain $d_\alpha$-dimensional homology class appears in the filtration at the moment $\tb_\alpha$, and vanishes at the moment $\td_\alpha$ (if the homology does not vanish, then $\td_\alpha$ is set to $+\infty$). The difference $\td_\alpha-\tb_\alpha$ is called the lifetime of the persistent homology. Usually, persistent homologies are visualized by persistent diagrams (PDs): each pair is represented by a point $(\tb_\alpha,\td_\alpha)$ on the plane $\Ro\times(\Ro\sqcup\{+\infty\})$, while the dimension $d_\alpha$ of the corresponding homology is visualized by a certain color.

In TDA, persistent homology allows to distinguish important topological features of filtrations from noise in the data. If some homology $\alpha$ has large lifetime (i.e. the corresponding point $(\tb_\alpha,\td_\alpha)$ is far from the diagonal $\td=\tb$), this means that this homology is important. The theory of Wasserstein--Kantorovich \cite{WK} metrics is developed in order to make such quantitative analysis more strict.
\end{con}

Returning to place cells' activity, we build the filtration $\{K_\theta\mid \theta\in[0;1]\}$ depending on the threshold parameter $\theta$. By computing the PH of this filtration, it is possible to discover the topology of the environment. If the physical space $X$ has $k$ obstacles, i.e. $\beta_1(X)=k$, then we expect to observe $k$ points far from the diagonal on the persistence diagram of 1-dimensional homology of $\{K_\theta\}$.


More constructively: assume that there are $m$ place cells, the activity of which depends only on the point in the physical space $X$ where the animal is located (and independent from time). Now assume that each neuron's activity is not binary, but takes real values from 0 to 1. Let the function $\phi_i\colon X\to \Ro$ denote the activity of $i$-th neuron at points of $X$. We assume that all $\phi_i$'s are continuous. Roughly speaking, the place field $U_i$ is the support set $\supp \phi_i$. All activity functions can be combined into a single continuous map to the $m$-dimensional space
\[
\Phi\colon X\to \Ro^m,\quad \Phi(x)=(\phi_1(x),\ldots,\phi_m(x))
\] 
Let us also assume that $\Phi$ distinguishes points of $X$, that is, for each pair of distinct points $x_1\neq x_2$ of $X$ at least one neuron $i$ has different activities at these points: $\phi_i(x_1)\neq\phi_i(x_2)$. Then, since $X$ is compact and $\Ro^m$ is Hausdorff, the image $\Phi(X)\subset \Ro^m$ is homeomorphic to $X$.

Now, while the animal moves in the environment, a finite number of measurements is taken at discrete moments of time $1,\ldots,T$. Each measurement of place cells' activity gives a point $y_t\in \Phi(X)$. So far, our measurements produce a point cloud $Y=\{y_t\mid 1\leqslant t\leqslant T\}$ sampled from the image $\Phi(X)$. One can try to estimate the topology of this point cloud using TDA methods. The standard idea here is to construct the Vietoris--Rips filtration $\{K_\theta^{VR}\}$ or witness complexes \cite{witt} of the point cloud and compute its persistent homology. However, if $m$ is large enough (say, $>10$), it might be a good idea to reduce the dimension as usually most valuable information lies in a lesser-dimensional space. It also would help to filter out the unnecessary noise. So, at the first step, it is natural to reduce the ambient dimension of the point cloud using some standard methods, e.g. PCA, Isomap \cite{iso}, t-sne, etc.. After the dimension reduction one will be able to compute persistent homology, and find the number of persistent 1-cycles, corresponding to the number of obstacles in $X$.

The point cloud model seems more preferable than the neural code model for several reasons. First, the assumption that activity is a function is more realistic. Second, the neural code model is sensitive to contractibility condition required by the Nerve theorem. In the point cloud model we do not require that supports of functions and their intersections are contractible. In case of the actual data, the activity patterns might be disconnected as well as non-simply connected, which is the reason why the point cloud model might be better suited for real experiments.

\subsection{Relation between the two models}
In this paragraph we relate the two models -- the nerve filtration and the point cloud filtration -- to each other. The main difficulty is that in the nerve model the simplicial complexes are defined on the set $[m]$ of neurons, while in the point cloud model the simplicial complexes are defined on the set $[T]$ of all time moments. To relate these two settings we use Dowker duality.  

Let us assume that the activity of $i$-th neuron is modeled by a function $\phi_i\colon X\to [0;1]$. For each $\theta\in[0;1]$ consider the superlevel sets $U_i^\theta=\{x\in X\mid \varphi_i(x)\geqslant 1-\theta\}$ and the nerves $K_\theta$ of the coverings $\bigcup_{i\in[m]}U_i^\theta$. The complexes $\{K_\theta\}_{\theta\in [0;1]}$ form an increasing filtration: if $\theta_1<\theta_2$, then $U_i^{\theta_1}\subseteq U_i^{\theta_2}$ and therefore $K_{\theta_1}\subseteq K_{\theta_2}$. According to the Nerve theorem, the PH of this filtration represents the topology of $X$ if all subsets $U_i^\theta$ are convex. This motivates the following definition, due to Itskov \cite{Itskov2}:

\begin{defin}
The function $\phi\colon X\to \Ro$ is called upper convex, if every superlevel set $\{x\in X\mid \phi(x)\geqslant c\}$ is convex.
\end{defin}

To make a bridge between the two models of neural activity described above, we assume that all neural activity functions $\phi_i$ are upper convex, i.e. $U_i^\theta$ is convex for all $i\in[m]$ and $\theta\in[0;1)$ (when $\theta=1$, one has $U_i^1=X$: the whole space is not required to be convex). With all neural activities $\phi_i$ being fixed, we obtain a continuous map $\Phi\colon X\to [0;1]^m$ as in the previous paragraph. 

Assume that a finite set $Y=\{y_1,\ldots,y_T\}\subset X$ is fixed, where $y_t$ is the position of the animal at moment of time $t$. The neural activities' measurements are given by a $T\times m$ table $A\colon [T]\times [m]\to [0;1]$, $A(t,i)=\phi_i(y_t)$. With the threshold parameter $\theta$ fixed, we get a binary table $A_\theta\colon [T]\times [m]\to \{0,1\}$,
\[
A_\theta(t,i)=\begin{cases}
1, & \mbox{if } A(t,i)\geqslant 1-\theta; \\
0, & \mbox{otherwise}.
\end{cases}
\]
We assume that the set $Y$ is dense enough in $X$ -- that is, whenever a collection $\{U_{i}^\theta\}_{i\in S}$ of subsets intersect, the intersection contains at least one point from $Y$. This assumption guarantees that the topology of $X$ can be recovered from the metric properties of the finite set $Y$. 

\begin{con}
Let $B\colon [T]\times [m]\to \{0,1\}$ be a binary table (which can be considered as a Boolean matrix, or a relation between $[T]$ and $[m]$, i.e. a subset of $[T]\times[m]$). A simplicial complex, called the Dowker complex, can be constructed from $B$. By definition $\Dowk(B)$ is a simplicial complex on the vertex set $[m]$, and $I\subseteq[m]$ is a simplex of $\Dowk(B)$ if and only if there exists $t\in [T]$ such that $B(t,i)=1$ for each $i\in I$. In other words, simplices of $\Dowk(B)$ are generated by the rows of the matrix $B$. Symmetrically, one can generate a complex by the columns of the matrix, corresponding to the transposed matrix: $\Dowk(B^\top)$. The famous result of Dowker asserts the homotopy equivalence $\Dowk(B^\top)\simeq \Dowk(B)$. Historically, Dowker proved~\cite{Dowker} the isomorphism of homology groups $H_*(\Dowk(B^\top))\cong H_*(\Dowk(B))$, while the homotopy result seem to first appear in \cite{Dowker}. In the recent paper~\cite{ChowMe} it was proved that homotopy equivalence of Dowker complexes respects functoriality: if $B_1\subset B_2$ are two binary relations, then the homotopy equivalences $\Dowk(B_1^\top)\simeq\Dowk(B_1)$ and $\Dowk(B_2^\top)\simeq\Dowk(B_2)$ can be chosen to commute with the natural maps $\Dowk(B_1)\hookrightarrow \Dowk(B_2)$ and $\Dowk(B_1^\top)\hookrightarrow \Dowk(B_2^\top)$. This observation allows to extend Dowker duality to filtrations of binary relations.
\end{con}

For the binary relation $A_\theta$ it holds true that $\Dowk(A_\theta)=K_\theta$ when $Y\subset X$ is dense enough. Indeed, $I=\{i_1,\ldots,i_s\}\in \Dowk(A_\theta)$ whenever there exists a time moment $t$ such that the neurons $i_1,\ldots,i_s$ are active at the point $y_t\in Y\subset X$. From the Dowker theorem it follows that

\begin{prop}
The nerve complex $K_\theta$ is homotopy equivalent to the $\Dowk(A_\theta^\top)$ complex on the vertex set $[T]$ where $\{t_1,\ldots,t_s\}$ is a simplex if there exists $i\in [m]$ such that $\phi_i(t_1),\ldots,\phi_i(t_s)\geqslant 1-\theta$.
\end{prop}

The filtration $\{\Dowk(A_\theta^\top)\}$ can be reconstructed from the point cloud $\Phi(Y)\subset\Ro^m$ as follows. A simplex $J=\{t_1,\ldots,t_s\}$ appears in the filtration $\{\Dowk(A_\theta^\top)\}_\theta$ at moment 
\[
\theta_J^{nerve}=\min_{i\in[m]}(1-\min(\phi_i(t_1),\ldots,\phi_i(t_s))).
\]
Recall that, in the point cloud model, the simplex $J=\{t_1,\ldots,t_s\}$ appears in the Vietoris--Rips filtration $\{K^{VR}_\theta\}$ at moment
\[
\theta_J^{cloud}=\max_{i<j}\dist(\Phi(t_i),\Phi(t_j)).
\]

\section*{Acknowledgments}

The article was prepared within the framework of the HSE University Basic Research Program.

This research was supported in part through computational resources of HPC facilities at HSE University \cite{charisma}.

%
%
%

\end{document}